\documentstyle [twoside,12pt]{article}
\newcommand{\nc}{\newcommand}
\nc{\la}{\lambda} \nc{\al}{\alpha}
\nc{\th}{\theta}  \nc{\be}{\beta}
\nc{\ga}{\gamma}  \nc{\Ga}{\Gamma}
\nc{\de}{\delta} \nc{\De}{\Delta}
\nc{\si}{\sigma}  \nc{\ka}{\kappa}
\nc{\om}{\omega}
\nc{\ra}{\rightarrow}
\nc{\beq}{\begin{equation}}
\nc{\eeq}{\end{equation}}
\nc{\beqa}{\begin{eqnarray}}
\nc{\eeqa}{\end{eqnarray}} \nc{\nnb}{\nonumber}
\nc{\dst}{\displaystyle}
\title{ {\bf B.R.S. renormalisation of some on-shell closed algebras of
symmetry transformations :} \\ 2) N=2 and 4 supersymmetric non-linear $\si$
models  }
\author{Guy Bonneau
\thanks{\noindent Laboratoire de Physique Th\'eorique et des Hautes Energies,
 Unit\'e associ\'ee au CNRS URA 280,~Universit\'e Paris 7,
 2 Place Jussieu, 75251 Paris Cedex 05. Electronic adress
bonneau@lpthe.jussieu.fr} }
\begin{document}
\maketitle
\begin{abstract}

\noindent We analyse with the algebraic, regularisation independant,
cohomological B.R.S. methods, the renormalisability of torsionless N=2 and N= 4
supersymmetric non-linear $\si$ models built on K\"ahler spaces. Surprisingly
enough with respect to the common wisdom, in the case of N=2 supersymmetry, we
obtain an anomaly candidate, at least in the compact K\"ahler Ricci-flat case.
If its coefficient does differ from zero, such anomaly would imply the breaking
of global N=2 supersymmetry and get into trouble some schemes of superstring
compactification as such non-linear $\si$ models offer candidates for the
superstring vacuum state.

In the compact homogeneous K\"ahler case, as expected, the anomaly candidate
disappears.

The same phenomena occurs when one enforces N=4 supersymmetry : in that case,
we obtain the first rigorous proof of the expected all-orders renormalisability
-`` in the space of metrics"- of the corresponding non-linear $\si$ models.
\end{abstract}

\vfill
{\bf PAR/LPTHE/94-11}\hfill  Mai 1994

\section{Introduction}

Supersymmetric non-linear $\si$  models in two space time dimensions have been
considered for many years to describe the vacuum state of superstrings
\cite{1},\cite{GS1}. In particular Calabi-Yau spaces, {\it i.e.} 6 dimensional
compact K\"ahler Ricci-flat Riemanian manifolds \cite {2}, appear as good
candidates in the compactification of the 10 dimensional superstring to 4
dimensional flat Minkowski space; indeed, the conformal invariance of the 2.d,
N = 2 supersymmetric non-linear $\si$  model (the fields of which are
coordinates on this compact manifold) is expected to hold to all orders of
perturbation theory \cite {3}.

However explicit calculations to 4 or 5 loops \cite {4} and, afterwards,
general arguments \cite{5} show that the $\be$ functions might be different
from zero.
But, as argued in my recent review \cite{6}, at least two problems obscure
these analyses : first, the fact that the quantum theory is not sufficiently
defined by the K\"ahler Ricci-flatness requirement ; second, the use of
``dimensional reduction" \cite{s} or of harmonic superspace \cite{r1}
\footnote{\ The regularisation through dimensional reduction suffers from
algebric unconsistencies and the quantisation in harmonic superspace does not
rely on firm basis, due to the presence of non-local singularities (in the
harmonic superspace)\cite{r2}.} in actual explicit calculations and general
arguments. Then, we prefer to analyse these models using the B.R.S., algebric,
regularization free cohomological methods.

Moreover, the quantisation of extended supersymmetry raises the difficulty of
an ``on-shell" formalism. Indeed, if one leaves aside harmonic superspace
where firm rules for quantisation \footnote{\  {\sl i.e.} a subtraction
algorithm insuring the locality of the counterterms \cite{r2}.} are not at
hand, on the contrary of ordinary superspace \cite{10}, one has to deal with
(super)symmetry transformations that are non-linear and closes only on-shell.
This problem was adressed in ref. \cite{ps1} by O. Piguet and K. Sibold for the
Wess-Zumino model as a ``toy-model" and, in a still uncomplete way, by P.
Breitenlohner and D. Maison \cite{bm} for supersymmetric Yang-Mills in the
Wess-Zumino gauge ; in the first paper of this series \cite{bo1}, hereafter
refered to as (I), we analysed the d=2, N=1 supersymmetric non-linear $\si $
model without auxiliary fields.

In the second paper of this series, we adress ourselves the question of the
all-orders renormalisability of extended supersymmetric (N = 2, 4) non-linear
$\si$   models in two space time dimensions. Of course, we are only interested
here in the renormalisation of the supersymmetry transformations : as discussed
by Friedan \cite{100}, the action of a non-linear $\si$ model may be identified
with a distance on a Riemannian manifold $\cal{M}$, the metric depending {\sl a
priori} on an infinite number of parameters. One then speaks of
``renormalisability in the space of metrics" or ``{\sl \`a la } Friedan". When
there exist extra  isometries, for example in the case of the non-linear $\si$
models on coset spaces (homogeneous manifolds), the number of such physical
parameters becomes finite and we have proved the U.V. renormalisability of
these isometries in the purely bosonic case in \cite{7}, as well as in the N=1
supersymmetric extension in (I). The present work gives the necessary
extended-supersymmetric generalisations. On the other hand, in the generalised
non-linear $\si$ models {\sl \`a la } Friedan, our aim is the proof that no
extra difficulty occurs in their supersymmetric extension.

We shall use N=1 superfields, which is allowed by the general superspace
quantisation methods established by Piguet and Rouet who in particular
demonstrated the Quantum Action Principle in that context \cite{10}, and the
very results of (I), proving that N=1 supersymmetry is all-orders
renormalisable. The classical theory was defined in (I), so here we only recall
in subsection 2.1 the results needed for the following. Due to the
non-linearity of the supersymmetry transformations in a general field
parametrisation ( $i.e.$ coordinate system on the manifold), we shall use a
gradation (according to the spectral sequences method \cite{14}) in the number
of fields, ghosts and their derivatives. As a matter of facts, we find it
convenient to use  two successive gradations, one in the number of extra
supersymmetries, the second one with respect to the number of fields. The
``filtrations" , as well as the lowest order nihilpotent Slavnov operators:
$S_L^{0}$ - corresponding in fact to N=2 supersymmetry -, and $S_L^{00}$ -
corresponding to the zero field approximation of $S_L^{0}$ -, are defined in
subsection 2.2. As in \cite{7} and (I), the cohomology of  $S_L^{00}$ will give
the main information.
In Section 3, we analyse the cohomology of $S_L^{00}$ and in Section 4 the one
of $S_L^0$, $i.e.$ at that point we are concerned with the special case of N=2
supersymmetric non-linear $\si$ models, and we find a non trivial cohomology in
the anomaly sector. Subsection 4.4 is then devoted to a discussion of this N=2
case and our main result is that, surprisingly enough with respect to the
common wisdom \cite{18}\footnote{\ Notice also that recent works of Brandt
\cite{8} and Dixon \cite{9} show the existence of new non-trivial cohomologies
in supersymmetric theories.}, there exists a possible anomaly for
\underline{global} supersymmetry in 2 space-time dimensions \cite{bo2}, at
least for \underline{torsionless} compact K\"ahler Ricci-flat manifolds ({\it
i.e.} special N=2 supersymmetric models). We also prove that this anomaly
disappears when the manifold $\cal{M}$ is an homogeneous one, $i.e.$ when one
deals with N=2 supersymmetric non-linear $\si$ models on coset spaces.
Section 5 then constructs the cohomology space of the complete $S_L$ operator,
with the essential result of the all orders renormalisability of N=4
supersymmetric non-linear $\si$ models. A discussion of our results is
presented in the concluding Section.

\section{The classical theory and the Slavnov operator}
In (I) we obtained the classical action and the linearised Slavnov operator
that describes N=4 supersymmetry and hereafter we summarize the essential
results.

\subsection{The classical theory and the Slavnov identity}
We consider d=2, N=4 supersymmetric non-linear $\si$ models in N=1 superfields
$\Phi^{i} (x,\th)$ (i, j,.. = 1,2,..4n). In light-cone coordinates and
\underline{in the absence of torsion}, the non-linear N=4 supersymmetry
transformations write :
\beq\label{C1}
\de\Phi^i = J_{A\,j}^{i}(\Phi)[\epsilon^{+}_A D_{+}\Phi^j + \epsilon^{-}_A
D_{-}\Phi^j ] \ \ ,\ \ A = 1, 2, 3.
\eeq
where the covariant derivatives are
 $$ D_{\pm} = \frac{\partial}{\partial\th^{\pm}} + i \th^{\pm}\partial_{\pm} $$
and satisfy
\beq\label{C2}
\{ D_{\pm},D_{\pm} \} = 2i\partial_{\pm} \ \ \ \ \{ D_{+},D_{-} \} = 0
\eeq

As is well known (see for example ref.\cite{11}), N=4 supersymmetry needs the
$J_{A\,j}^{i}(\Phi)$ to be a set \footnote{\ As a matter of facts, it is
sufficient to have 2 anticommuting integrable complex structures : then, the
product  $J_{3\,k}^{i} \equiv J_{1\,j}^{i} J_{2\,k}^{j}$ offers a third complex
structure.} of anticommuting integrable complex structures according to :
\beq\label{C3}
J_{A\,j}^{i}(\Phi) J_{B\,k}^{j}(\Phi) = -\de_{AB}\de^i_k +
\epsilon_{ABC}J_{C\,k}^{i}(\Phi)
\eeq
and the invariance of the action $A^{inv.} = \int d^2x d^2\th
g_{ij}[\Phi]D_+\Phi^iD_-\Phi^j$  needs the target space to be hyperk\"{a}hler :

$\ast$ the metric $g_{ij}$ is hermitic with respect to each complex structure
$$J_{A}^{ij} \equiv J_{A\,k}^{i}g^{kj} = - J_{A}^{ji}\ \ ;\ \ J_{A\,ij} \equiv
J_{A\,i}^{k}g_{kj} = - J_{A\,ji}$$

$\ast$ the $J_{A\,j}^{i}$ are covariantly constant
$$D_{k}J_{A\,j}^{i} \equiv \partial_k J_{A\,j}^{i} +\Ga^i_{kl}J_{A\,j}^{l} -
\Ga^l_{kj}J_{A\,l}^{i} = 0 $$
where $\Ga^i_{kl}$ is the (symmetric) connexion  with respect to the metric
$g_{ij}$.
In the B.R.S. approach \cite{bp}, the supersymmetry parameters
$\epsilon_A^{\pm}$ are promoted to constant, commuting Faddeev-Popov parameters
$d_A^{\pm}$ \footnote{\ As one is only concerned by integrated local
functionals - $i.e.$ trivially translation invariant ones -, we forget about
the linear translation operators $P_\pm \equiv  i\partial_\pm $, to which
anticommuting Faddeev-Popov parameters $p^\pm $ should be associated, and do
not add in $\Ga^{class.}$ of equ.(\ref{C4}) the effect of translations on the
fields $\Phi^i$ .}  and an anticommuting classical source $\eta_i(x)$  for the
non-linear field transformation (\ref{C1}) is introduced in the classical
action \footnote {\ In the absence of torsion, there is a parity invariance $$
+ \ra -, d^2x \ra d^2x, d^2\th \ra -d^2\th, \Phi^i \ra \Phi^i, \eta_i \ra
-\eta_i \ .$$ Moreover, the canonical dimensions of $[ d^2x d^2\th],\
[\Phi^i],\ [d_A^{\pm}],\ [D_{\pm}],\ [\eta_i]$ are -1, 0, -1/2, +1/2, +1
respectively and the Faddeev-Popov assignments + 1 for $d_A^{\pm}$, -1 for
$\eta_i$, 0 for the other quantities.}. Then, the total effective action
\footnote{\ For simplicity, no mass term has been added here as here we are
only interested in U.V. properties .}  is :
\beq\label{C4}
\Ga^{class.} = A^{inv.} + \int d^2 x d^2 \th \{ \eta_i J_{A\,j}^{i}(\Phi)[
d_A^{+}D_{+}\Phi^j + d_A^{-}D_{-}\Phi^j ] - {1\over2}\epsilon_{ABC}\eta_i\eta_j
J_{C}^{ij}(\Phi)d_A^{+}d_B^{-} \}
\eeq
The terms quadratic in the sources are needed as a consequence of the only
on-shell closedness of the N=4 supersymmetry algebra \cite{bv}(I).

The Slavnov identity writes :
\beq\label{C5}
S \Ga^{class.} \equiv \int d^2 x d^2 \th\frac{\de \Ga^{tot.}}{\de
\eta_i(x,\th)} \frac{\de \Ga^{tot.}}{\de \Phi^i(x,\th)} = \int d^2 x d^2 \th
[(d_A^+)^2 ( \eta_k i\partial_+\Phi^k)  + (d_A^-)^2 ( \eta_k i\partial_-\Phi^k)
]\ .
\eeq
This is a not trivial result as in that N = 4 case, no finite set of auxiliary
fields does exist.

As is by now well known (for example see \cite{6} or \cite{7}), in the absence
of a consistent regularisation that respects all the symmetries of the theory,
the quantum analysis directly depends on the cohomology of the nihilpotent
linearised Slavnov operator :
\beqa\label{C6}
S_L &=& \int d^2x d^2\th \left[\frac {\de\Ga^{class.}}{\de\eta_i(x,\th)}\frac
{\de}{\de\Phi^i(x,\th)} + \frac {\de\Ga^{class.}}{\de\Phi^i(x,\th)}\frac
{\de}{\de\eta_i(x,\th)} \right] \nnb\\
S_L^2 &=& 0
\eeqa
in the Faddeev-Popov charge +1 sector [absence of anomalies for the
supersymmetry] and 0 sector [number of physical parameters and stability of the
classical action through radiative corrections]. Notice that the Slavnov
operator (\ref{C6}) is unchanged under the following field and source
reparametrisations :
$$\Phi^i \ra \Phi^i + \la W^i[\Phi] \ \ \ ,\ \ \eta_i \ra \eta_i - \la \eta_k
W^k_{,i}[\Phi] \ ,$$
where $W^i[\Phi]$ is an arbitrary function of the fields $\Phi(x,\th)$ and a
comma indicates a derivative with respect to the field $\Phi^i$. Under this
change, the classical action (\ref{C4}) is modified :
\beq\label{C7}
\Ga^{class.} \ra \Ga^{class.} + \la S_L \int d^2x d^2\th \eta_i W^i[\Phi]
\eeq
but the Slavnov identity is left unchanged as
\beq\label{C71}
S[\Ga^{class.} + \la S_L\De] \equiv S\Ga^{class.} + \la S_L[S_L \De] =
S\Ga^{class.} \ .
\eeq

\noindent The quantisation of this theory will be studied in the next Sections,
using the same algebraic cohomological methods as in the first paper of this
series (I). It will be convenient to separate the 3 extra supersymmetries into
the one \footnote{\ In the following, we then omit the index 3 of the complex
struture $J_3$ as well as the one of the ghost $d_3^{\pm}$ .} corresponding to
$J_3$ and the 2 others to $J_{\al},\ \al$=1, 2, $i.e.$ to separate the N=2
supersymmetric case from the N=4 one. In the same way, one splits the
linearised Slavnov operator into 3 parts according to their number of ghosts
$d^{\pm}_{\al}$ :
$$S_L = S_L^0 + S_L^1 + S_L^2$$
$$(S_L^0)^2 = S_L^0 S_L^1 + S_L^1 S_L^0 = S_L^0 S_L^2 +  S_L^1 S_L^1 + S_L^2
S_L^0  = S_L^1 S_L^2 + S_L^2 S_L^1 = (S_L^2)^2 = 0 \ .$$
\beqa\label{C8}
S_L^0 &=&  \int d^2x d^2\th\{J^i_j (d^+ D_+\Phi^j + d^- D_-\Phi^j)
\frac{\de}{\de\Phi^i} \nnb\\
&+& [ \frac{\de A^{inv.}}{\de \Phi^i} +\eta_k (J^k_{j\,,i}-J^k_{i\,,j})(d^+
D_+\Phi^j + d^- D_-\Phi^j) + J^j_i(d^+ D_+\eta_j + d^- D_-\eta_j)]
\frac{\de}{\de\eta_i}\}\ ,
\eeqa
which does not change the number of ghosts $d^{\pm}_{\al}$, will play a special
role. Moreover, notice that the cohomology of $S_L^{0}$ corresponds to the
special N=2 supersymmetric case.

\subsection{The filtration and the operator $S_L^0$}

In the presence of highly non-linear Slavnov operators such as in (\ref{C8}),
as recalled in (I), it is technically useful to ``aproximate" the complete
$S_L^0$ operator by a simpler one $S_L^{00}$ through a suitably chosen
``filtration"( ghost number preserving counting operation)\cite{14}. As it does
not change this number, $S_L^{00}$, the nihilpotent lowest order part of
$S_L^0$, will play a special role. Here, we take as counting operator the total
number of fields $\Phi^i(x,\th)$ and their derivatives. Then :
\beqa\label{a5}
S_L^0 &=& S_L^{00} + S_L^{01} + S_L^{02} +... \equiv S_L^{00} + S_L^r \ ,\ \
(S_L^{00})^2 = 0 \nnb\\
S_L^{00} &=&  J^i_j(0)\int d^2x d^2\th\left\{(d^+ D_+\Phi^j + d^- D_-\Phi^j)
\frac{\de}{\de\Phi^i} +
(d^+ D_+\eta_i + d^- D_-\eta_i)\frac{\de}{\de\eta_j}\right\} \ .
\eeqa

\noindent As explained in refs.(\cite{14},\cite{7}), when $S_L^{00}$ has no
cohomology in the Faddeev-Popov charged sectors, the cohomology of the complete
$S_L^0$ operator in the Faddeev-Popov sectors of charge 0 and +1 is isomorphic
to the one of  $S_L^{00}$ in the same sectors. The extension to the case where
$S_L^{00}$ has some non-trivial cohomology was discussed in the appendix of
(I)\footnote {\ In particular, the cohomology of $S_L^{00}$ in the
Faddeev-Popov -1 sector restricts the dimension of the cohomology of $S_L^0$ in
the 0 charge sector when compared to the one of  $S_L^{00}$.} (see also the
original papers \cite{14},\cite{7} and \cite{bp2}).

Then, in the next Section, we shall determine the cohomology spaces of
$S_L^{00}$ in the Fadeev-Popov sectors of charge -1, 0 and +1.

\section{The cohomology of $S_{L}^{00}$}

The most general functional in the fields, sources, ghosts and their
derivatives, of a given Faddeev-Popov charge, is built using Lorentz and parity
invariance and power counting (see footnote 6).

\subsection{The Faddeev-Popov negatively charged sectors}
Due to dimensions and Faddeev-Popov charge assignments, dimension zero
integrated local polynomials in the Faddeev-Popov parameters, fields, sources
and their derivatives have at least a Faddeev-Popov charge -1 :
\beq\label{a7}
\De_{[-1]} = \int d^2x d^2\th \eta_iV^i[\Phi]\ .
\eeq
Then there is no Faddeev-Popov charge -1 coboundaries, so the cohomology of
$S_L^{00}$ in that sector is given by the cocycle condition :
\beq\label{a8}
S_L^{00} \De_{[-1]} = 0 \ \ \Leftrightarrow \ \ J^i_j(0)V^k_{,i} =
J^k_i(0)V^i_{,j}
\eeq
This condition, when expressed in a coordinate system adapted to the complex
structure  $J^i_j[\Phi]$ (i $\equiv (a,\ \bar{a}),\ \Phi^i \equiv (\phi^a,\
\bar{\phi}^{\bar{a}})$ : $J^{a}_{b} = i \de^{a}_{b} , J^{\bar{a}}_{\bar{b}} =
-i \de^{a}_{b}, J^{\bar{a}}_{b} = J^{a}_{\bar{b}} = 0 $), means that
$V^i[\Phi]$ is a contravariant analytic vector : $V^{a} = V^{a}[\phi^{d}] ,\
V^{\bar{a}} = V^{\bar{a}}[\bar{\phi}^{\bar{d}}]$.

Let us now turn to the Faddeev-Popov neutral charge sector.

\subsection{The Faddeev-Popov 0 charge sector }

Here, one decomposes the set of integrated local polynomials in the
Faddeev-Popov parameters, fields, sources and their derivatives with respect to
their number of ghosts $d^{\pm}_{\al},\ N_{d_{\al}}\ .$

\beqa\label{a81}
\De_{[0]}^0 &=& \int d^2x d^2\th \left\{ t_{ij}[\Phi]D_+\Phi^iD_-\Phi^j +
\eta_i U^i_j[\Phi](d^+ D_+\Phi^j + d^- D_-\Phi^j) \right\}\nnb\\
\De_{[0]}^1 &=& \int d^2x d^2\th \left\{ \eta_i U^i_{\al\,j}[\Phi](d_{\al}^+
D_+\Phi^j + d_{\al}^- D_-\Phi^j) +\eta_i\eta_j(d^+_{\al}d^- -
d^-_{\al}d^+)S_{\al}^{ij}[\Phi]\right\}\nnb\\
\De_{[0]}^2 &=&  d_{\al}^+ d_{\be}^-\int d^2x d^2\th
\eta_i\eta_jS_{[\al\be]}^{ij}[\Phi]
\eeqa
where, due to parity invariance (footnote 6), $t_{ij}\ $(resp. $S_{\al}^{ij},\
S_{[\al\be]}^{ij})$ are symmetric (resp. skew-symmetric) in (i,j). Coboundaries
being given by $S_L^{00} \De_{[-1]}$[arbitrary $V^i(\Phi)$], the analysis of
the cocycle condition $S_L^{00} \De_{[0]} = 0 $ successively gives :

\subsubsection{$N_{d_{\al}}=0$}
\beq\label{a811}
\De_{[0]}^0 =  \De_{[0]}^{an.}[t_{ij}(\Phi)] + S_L^{00} \De_{[-1]}[V^i(\Phi)]
\ \ ;\ \De_{[0]}^{an.}[t_{ij}] = \int d^2x d^2\th
t_{ij}[\Phi]D_+\Phi^iD_-\Phi^j
\eeq
where the tensor  $t_{ij}$ which occurs in the anomalous part is constrained by
:
\beqa\label{a9}
a) & \ \ J^i_j(0)t_{ik} + t_{ji}J^i_k(0) & = 0 ,\nnb \\
b) & \ \ J^i_j(0)[t_{kl,i}-t_{il,k}] - (j \leftrightarrow k ) &= 0.
\eeqa

\noindent The absence of source dependent non-trivial cohomology means that, up
to a field redefinition (see (\ref{C7},\ref{C71})), the complex structure
$J_j^i$ is left unchanged through radiative corrections. Moreover, condition
(\ref{a9}a)   means that the metric $g_{ij} + \hbar t_{ij}$ remains hermitian
with respect to the complex structure $J_j^i$, whereas (\ref{a9}b) expresses
the covariant constancy of $J^i_j$ with respect  to the covariant derivative
with a connexion corresponding to the metric $g_{ij} + \hbar t_{ij}$. These are
precisely the expected conditions for the stability of N=2 supersymmetry.

\subsubsection{$N_{d_{\al}}=1$}
$S_L^{00} \De_{[0]}^1 = 0 $ gives (no coboundaries exist in that sector) :
\beq\label{a812}
U^i_{\al\,j}=0\ \ ,\ \ \De_{[0]}^1 =  \De_{[0]}^{an.}[S_{\al}^{ij}(\Phi)] =
\int d^2x d^2\th \eta_i\eta_j(d^+_{\al}d^- - d^-_{\al}d^+)S_{\al}^{ij}[\Phi]\ \
,
\eeq
where the tensor  $S_{\al}^{ij}$ which occurs in the anomalous part is
constrained by :
\beqa\label{a91}
a) & \ \ J^i_k(0)S_{\al}^{kj} + S_{\al}^{ki}J^j_k(0) &= 0 ,\nnb \\
b) & \ \ J^n_k(0)S_{\al\,,n}^{ij} - J^i_n(0)S_{\al\,,k}^{nj} &= 0\ ,
\eeqa
$i.e.$, using the same adapted coordinate system as above, is a pure
contravariant analytic skew-symmetric tensor ({\it i.e.} $S_{\al}^{[ab]} =
S_{\al}^{[ab]}(\phi^{c})$ , $ S_{\al}^{[\bar{a}\bar{b}]}  =
S_{\al}^{[\bar{a}\bar{b}]}(\bar{\phi}^{\bar{c}})$ , the other components
vanish).
\subsubsection{$N_{d_{\al}}=2$}
$S_L^{00} \De_{[0]}^2 = 0 $ gives (no coboundaries exist in that sector) :
\beq\label{a813}
\De_{[0]}^2 =  \De_{[0]}^{an.}[S_{[\al\be]}^{ij}(\Phi)] =
d_{\al}^+d^-_{\be}\int d^2x d^2\th \eta_i\eta_jS_{[\al\be]}^{ij}[\Phi]
\eeq
where the tensor  $S_{[\al\be]}^{ij}$ which occurs in the anomalous part is
constrained by :
\beqa\label{a92}
a) & J^i_k(0)S_{[\al\be]}^{kj} + S_{[\al\be]}^{ki}J^j_k(0) &= 0 ,\nnb\\
b) & \ \ J^n_k(0)S_{[\al\be]\,,n}^{ij} - J^i_n(0)S_{[\al\be]\,,k}^{nj} &= 0\ ,
\eeqa
$i.e.$, using the same adapted coordinate system as above, is a pure
contravariant analytic skew-symmetric tensor.

Finally, let us consider the Faddeev-Popov charge +1 sector.

\subsection{The Faddeev-Popov +1 charge sector}

Here also, one decomposes the set of integrated local polynomials in the
Faddeev-Popov parameters, fields, sources and their derivatives with respect to
their number of ghosts $d^{\pm}_{\al},\ N_{d_{\al}}\ .$

\subsubsection{ $N_{d_{\al}}$ = 0}
$\De_{[+1]}^0 $ depends on 8 tensors :
\beqa\label{b10}
\De_{[+1]}^0 &=& \int d^2x d^2\th \{(d^+)^2(d^-)^2\eta_i\eta_j\eta_k t^{[ijk]}
\nnb\\
&+& d^+d^-[\eta_i\eta_j t^{[ij]}_{1\; n}(d^+D_+\Phi^n - d^-D_-\Phi^n) + \eta_i
s^{(ij)}_1 (d^+D_+\eta_j - d^-D_-\eta_j)]\nnb\\
&+& d^+d^- \eta_k t^k_{2\; [ij]}D_+\Phi^iD_-\Phi^j  \nnb\\
&+& (d^+)^2\eta_k( t^k_{4\; [ij]}D_+\Phi^iD_+\Phi^j +
t^k_{5\;j}(D_+)^2\Phi^j)\nnb\\
&+& (d^-)^2\eta_k(t^k_{4\; [ij]}D_-\Phi^iD_-\Phi^j +
t^k_{5\;j}(D_-)^2\Phi^j)\nnb\\
&+& d^+(\tilde {t}_{[ij]n}D_+\Phi^iD_+\Phi^jD_-\Phi^n + s_{2\;
(ij)}D_-D_+\Phi^iD_+\Phi^j)\nnb\\
&-& d^-(\tilde {t}_{[ij]n}D_-\Phi^iD_-\Phi^jD_+\Phi^n + s_{2\;
(ij)}D_+D_-\Phi^iD_-\Phi^j)\}
\eeqa
where, due to the anticommuting properties of $\eta_i$ and $D_{\pm}\Phi^i$ and
to the integration by parts freedom, the tensors $t^{[ijk]}$, $t^{[ij]}_{1\;
n}$, $t^n_{2\; [ij]}$, $t^n_{4\; [ij]}$, $\tilde {t}_{[ij]n}$ are
skew-symmetric in i, j, k, and $s^{(ij)}_1$, $s_{2\; (ij)}$ symmetric in i, j.
Here and in the following, the symmetry (resp. antisymmetry) properties of the
involved tensors in the exchange i to j are indicated by parenthesis (ij)
(resp. brackets [ij]).

Coboundaries being given by $S_L^{00} \De_{[0]}^0$[arbitrary ($t_{ij}[\Phi],
U^i_j[\Phi])]$, the analysis of the cocycle condition $S_L^{00} \De_{[+1]}^0 =
0$ leads to:
\beq\label{b91}
\De_{[+1]}^{an.(0)} = \int d^2x d^2\th
t^{[ijk]}(\Phi)(d^+)^2(d^-)^2\eta_i\eta_j\eta_k
\eeq
where the skew-symmetric tensor $t^{[ijk]}(\Phi)$ which occurs in the anomalous
part is constrained by:
\beqa\label{b11}
a)& \ \ J^i_n(0)t^{[njk]} \ \ \ {\rm is\ \ i,\ j,\ k\ \ skew-symmetric},\nnb\\
b)&  \ \ J^i_n(0)t^{[njk]}_{,m} = J^n_m(0)t^{[ijk]}_{,n}
\eeqa

\noindent Using the same adapted coordinate system as above, condition
(\ref{b11}a) means that the tensor  $t^{[ijk]}$ is a pure contravariant
skew-symmetric tensor ({\it i.e.} $t^{[abc]} , t^{[\bar{a}\bar{b}\bar{c}]} \neq
0$ , the other components vanish)
whereas (\ref{b11}b)    means that it is analytic ({\it i.e.} $t^{[abc]} =
t^{[abc]}(\phi^{d})$,  $t^{[\bar{a}\bar{b}\bar{c}]} =
t^{[\bar{a}\bar{b}\bar{c}]}(\bar{\phi}^{\bar{d}}))$. In particular, due to the
vanishing of $t^{[ab\bar{c}]}$,  such tensor cannot be a candidate for a
torsion tensor on a K\"ahler manifold \cite{151}.

As a first result, this proves that if the manifold $\cal M$ has a complex
dimension smaller than 3, there is no $N_{d_{\al}}$ = 0 anomaly candidate.

\subsubsection{ $N_{d_{\al}}$ = 1}

With an expansion similar to the one of $\De_{[+1]}^0,\ \De_{[+1]}^1$ now
depends on 11 tensors :
\beqa\label{b101}
\De_{[+1]}^1 &=& d_{\al}^+\int d^2x d^2\th \{
d^+(d^-)^2\eta_i\eta_j\eta_kt_{\al}^{[ijk]} \nnb\\
&+& d^+d^-[\eta_i\eta_j t^{[ij]}_{\al 1\; n}D_+\Phi^n  + \eta_iD_+\eta_j
s^{(ij)}_{\al 1}] +  d^-d^-[\eta_i\eta_j t'^{[ij]}_{\al 1\; n}D_-\Phi^n  +
\eta_iD_-\eta_j s'^{(ij)}_{\al 1}]\nnb\\
&+& d^- \eta_k[ t^k_{\al 2\; [ij]}D_+\Phi^iD_-\Phi^j + t^k_{\al 3\;
j}D_+D_-\Phi^j] \nnb\\
&+& d^+\eta_k[ t^k_{\al 4\; [ij]}D_+\Phi^iD_+\Phi^j + t^k_{\al
5\;j}(D_+)^2\Phi^j]\nnb\\
&+& [\tilde {t}_{\al[ij]n}D_+\Phi^iD_+\Phi^jD_-\Phi^n + s_{\al 2\;
(ij)}D_-D_+\Phi^iD_+\Phi^j] \}\nnb\\
&+& {\rm parity\ exchanged\ (according\ to\ footnote\ 6)}
\eeqa

\noindent Coboundaries being given by $S_L^{00}\De_{[0]}^1$[arbitrary ($
U^i_{\al\,j}[\Phi], S_{\al}^{ij}[\Phi])]$, the analysis of the cocycle
condition $S_L^{00} \De_{[+1]}^1 = 0$ leads to :
\beqa\label{b911}
\De_{[+1]}^{an.(1)} = \int d^2x d^2\th t_{\al}^{[ijk]}(\Phi)
[d_{\al}^+d^+(d^-)^2 + d_{\al}^-d^-(d^+)^2]\eta_i\eta_j\eta_k +\nnb\\
 +  \int d^2x d^2\th \tilde{t}_{\al\,[ij]k}(\Phi)
[d_{\al}^+D_+\Phi^iD_+\Phi^jD_-\Phi^k - d_{\al}^-D_-\Phi^iD_-\Phi^jD_+\Phi^k ]
\eeqa
where the constraints on the skew-symmetric tensors $t_{\al}^{[ijk]}(\Phi)$ and
$\tilde{t}_{\al\,[ij]k}(\Phi)$ which occur in the anomalous part are easily
solved in the same adapted coordinate system as above :

$\bullet$ the tensor  $t_{\al}^{[ijk]}$ is a pure contravariant analytic
skew-symmetric tensor,

$\bullet$ the tensor  $\tilde{t}_{\al\,[ab]\bar{c}} \equiv
\partial_{\bar{c}}[\partial_{a}t_{\al\,b}(\phi,\bar{\phi}) -
\partial_{b}t_{\al\,a}(\phi,\bar{\phi})]$ (and the complex conjugate relation),
the other components vanish.

\subsubsection{ $N_{d_{\al}}$ = 2}

Here, we separate in $\De_{[+1]}^2$ the terms symmetric in the exchange of the
indices $\al$ and $\be$ of the 2 ghosts $d_{\al}^{\pm}$ and $d_{\be}^{\pm}$
from the skew-symmetric ones :
\beqa\label{b102}
\De_{[+1]}^2|_{(\al\be)}& = & d_{\al}^+d_{\be}^+ \int d^2x d^2\th \{
(d^-)^2\eta_i\eta_j\eta_kt_{(\al\be)}^{[ijk]} +
d^-[\eta_i\eta_jD_+\Phi^kt^{[ij]}_{(\al\be)1\;k} +
\eta_iD_+\eta_js^{(ij)}_{(\al\be)1}]\nnb\\
&+& \eta_k[D_+\Phi^iD_+\Phi^jt_{(\al\be)4\,[ij]}^k +
(D_+)^2\Phi^jt_{(\al\be)5\,j}^k] \}+ {\rm parity\ exchanged} +\nnb\\
& + & d_{\al}^+d_{\be}^- \int d^2x d^2\th \{d^+d^-\eta_i\eta_j\eta_k
t'^{[ijk]}_{(\al\be)} + \nnb\\
& + &  [\eta_i\eta_j(d^+D_+\Phi^k - d^-D_-\Phi^k)t'^{[ij]}_{(\al\be)1\;k} +
\eta_i(d^+D_+\eta_j - d^-D_-\eta_j)s'^{(ij)}_{(\al\be)1}] +\nnb\\
& + & \eta_kD_+\Phi^iD_-\Phi^j t'^{k}_{(\al\be)4\,[ij]} \}\ \ \  ; \nnb\\
\De_{[+1]}^2|_{[\al\be]}& = & d_{\al}^+d_{\be}^- \int d^2x d^2\th \{
[\eta_i\eta_j(d^+D_+\Phi^n + d^-D_-\Phi^n)t^{[ij]}_{[\al\be]1\; n} + \eta_i
(d^+D_+\eta_j + d^-D_-\eta_j)s^{(ij)}_{[\al\be]1}]\nnb\\
& + & \eta_k[D_+\Phi^iD_-\Phi^jt^k_{[\al\be]4\; (ij)} +
D_+D_-\Phi^jt^k_{[\al\be]5\;j}] \}
\eeqa
Then, coboundaries being given by $S_L^{00}\De_{[0]}^2$[arbitrary $\
S_{[\al\be]}^{ij}(\Phi)]$, the analysis of the cocycle condition $S_L^{00}
\De_{[+1]}^2 = 0$ leads to :
\beqa\label{b912}
\De_{[+1]}^{an.(2)}& = & d_{\al}^+d_{\be}^+ \int d^2x d^2\th
\eta_i\left[(d^-)^2\eta_j\eta_kt_{(\al\be)}^{[ijk]}(\Phi) +
(D_+)^2\Phi^jt_{(\al\be)\, j}^i(\Phi)\right] + {\rm parity\ exchanged} +\nnb\\
& + & d_{\al}^+d_{\be}^-d^+d^- \int d^2x d^2\th
\eta_i\eta_j\eta_kt'^{[ijk]}_{(\al\be)}(\Phi) + \nnb\\
& + &  d_{\al}^+d_{\be}^- \int d^2x d^2\th \eta_i\left[D_+\Phi^jD_-\Phi^k
t_{[\al\be]\,(jk)}^{i}(\Phi) + D_+D_-\Phi^j t_{[\al\be]\,j}^{i}(\Phi)\right]
\eeqa
where the constraints on the tensors $t_{(\al\be)}^{[ijk]}(\Phi),\
t_{(\al\be)\, j}^i(\Phi),\ t'^{[ijk]}_{(\al\be)}(\Phi),\
t_{[\al\be]\,(jk)}^{i}(\Phi)$ and $t_{[\al\be]\,j}^{i}(\Phi)$ which occur in
the anomalous part are easily solved in the same adapted coordinate system as
above :

$\bullet$ the tensors  $t_{(\al\be)}^{[ijk]}(\Phi)$ and
$t'^{[ijk]}_{(\al\be)}(\Phi)$ are pure contravariant analytic skew-symmetric
tensors,

$\bullet$ the tensor  $t_{(\al\be)\, j}^i(\Phi)$ is a mixed analytic tensor,
({\it i.e.} $t_{(\al\be)\, b}^a = t_{(\al\be)\, b}^a(\phi^c)$, $t_{(\al\be)\,
\bar{b}}^{\bar{a}} = t_{(\al\be)\, \bar{b}}^{\bar{a}}(\bar{\phi}^{\bar{c}}))$,

$\bullet$ the tensor  $t_{[\al\be]\,(ab)}^{\bar{c}} \equiv
\partial_{a}\partial_{b}t_{[\al\be]}^{\bar{c}}(\phi,\bar{\phi})$ and the tensor
 $t_{[\al\be]\,b}^{\bar{c}} \equiv
\partial_{b}t_{[\al\be]}^{\bar{c}}(\phi,\bar{\phi})$ (and the complex conjugate
relations), the other components vanish.

\subsubsection{ $N_{d_{\al}}$ = 3}
In that sector, there are no coboundaries, and the analysis of the cocycle
condition $S_L^{00} \De_{[+1]}^3 = 0$ with :
\beqa\label{b103}
\De_{[+1]}^3 &=& d_{\al}^+d_{\be}^+d_{\ga}^-\int d^2x d^2\th \{
d^-\eta_i\eta_j\eta_kt_{(\al\be)\ga}^{[ijk]} + \eta_i\eta_j
t^{[ij]}_{(\al\be)\ga\,1\; k}D_+\Phi^k  + \eta_iD_+\eta_j
s^{(ij)}_{(\al\be)\ga\,1}\} +\nnb\\
& + & {\rm parity\ exchanged}
\eeqa
leads to :
\beq\label{b913}
\De_{[+1]}^{an.(3)} = d_{\al}^+d_{\be}^+d_{\ga}^-d^-\int d^2x d^2\th
\eta_i\eta_j\eta_k t_{(\al\be)\ga}^{[ijk]}(\Phi) + {\rm parity\ exchanged}
\eeq
where the constraints on the skew-symmetric tensor
$t_{(\al\be)\ga}^{[ijk]}(\Phi)$ which occurs in the anomalous part are easily
solved in the same adapted coordinate system as above and again means that it
is a pure contravariant analytic skew-symmetric tensor.

\subsubsection{ $N_{d_{\al}}$ = 4}

In that sector too, there are no coboundaries, and the analysis of the cocycle
condition\newline $S_L^{00} \De_{[+1]}^4 = 0$ with :
\beqa\label{b104}
\De_{[+1]}^4 &=& d_{\al}^+d_{\be}^+d_{\ga}^-d_{\de}^-\int d^2x d^2\th
\eta_i\eta_j\eta_kt_{((\al\be)(\ga\de))}^{[ijk]}
\eeqa
leads to :
\beq\label{b914}
\De_{[+1]}^{an.(4)} = d_{\al}^+d_{\be}^+d_{\ga}^-d_{\de}^-\int d^2x d^2\th
\eta_i\eta_j\eta_kt_{((\al\be)(\ga\de))}^{[ijk]}(\Phi)
\eeq
where the constraints on the skew-symmetric tensor
$t_{((\al\be)(\ga\de))}^{[ijk]}(\Phi)$ which occurs in the anomalous part are
easily solved in the same adapted coordinate system as above and again means
that it is a pure contravariant analytic skew-symmetric tensor.

This ends the analysis of the cohomology of $S_{L}^{00}$ and we are now in a
position to discuss the cohomology of the complete $S_L^0 \equiv S_L^{00} +
S_L^r$ operator.

\section{The cohomology of $S_{L}^0$}

It will be convenient to analyse the cohomology of $S_L^0$ in a coordinate
system where the complex structure $J_3$ is constant. $S_{L}^{00}$ of
equ.(\ref{a5}) is unchanged (with $J_j^i(0) \rightarrow J_j^i$, field
independent) and we note that
\beq\label{a55}
S_L^r = \int d^2x d^2\th \frac{\de
A^{inv.}}{\de\Phi^i(x,\th)}\frac{\de}{\de\eta_i(x,\th)}
\eeq
decreases the number of $\eta_i$ of one unity, whereas $S_{L}^{00}$ does not
change this number. Using this fact, we are able to construct the $S_L^0$
cohomology starting from the $S_L^{00}$ one : indeed,
\beq\label{a551}
S_L^0\De_{[\phi\pi]} \equiv (S_{L}^{00} + S_L^r)(\De_{[\phi\pi]}^{an.} +
\tilde{\De}_{[\phi\pi]}) = 0 \ \Leftrightarrow
S_{L}^{00}\tilde{\De}_{[\phi\pi]} = -S_L^r(\De_{[\phi\pi]}^{an.} +
\tilde{\De}_{[\phi\pi]})\ .
\eeq
Then, when ordered by decreasing order with respect to the total number of
$\eta_i$, the equation  $S_L^0\De_{[\phi\pi]} = 0$ is identical to the
$S_L^{00}\De_{[\phi\pi]} = 0$ one with a right hand side given by previous
order contributions.

\subsection{The Faddeev-Popov negatively charged sectors}

Thanks to the simplicity of $\De_{[-1]}$ (\ref{a7}), the cohomology of the
complete $S_L^0$ operator in the Faddeev-Popov charge -1 sector is easily
obtained : the vector $V^i[\Phi]$ should satisfy :
$$\bullet \ \  \int d^2x d^2\th \frac{\de A^{inv.}}{\de
\Phi^i(x,\th)}V^i[\Phi(x,\th)] = 0 \ \Leftrightarrow V^i[\Phi]\ {\rm is\  a\
Killing\ vector\ for\ the\ metric\ }\ g_{ij}[\Phi]$$
$$ \bullet J^i_j[\Phi]\nabla_iV^k = \nabla_jV^iJ^k_i[\Phi] \ \Leftrightarrow
V^i[\Phi]\ {\rm  is\ a\ contravariant\ vector\ analytic\ with\ respect\ to}\
J^i_j[\Phi]\ .$$
Let us now turn to the Faddeev-Popov neutral charge sector.

\subsection{The Faddeev-Popov 0 charge sector }

As explained in the appendix of (I) (see also \cite{7},\cite{14}), despite the
non-vanishing  $S_L^{00}$ cohomology in a Faddeev-Popov positively charged
sector (subsection 3.3), the cohomology of $S_L^0$ is a subspace of the one of
$S_L^{00}$, {\it i.e.} one can always construct the cocycles for $S_L^0$
starting from those of $S_L^{00}$.
It may also happen that some of the so doing constructed cocycles for $S_L^0$
become coboundaries: this occurs when there is some cohomology for  $S_L^{00}$
in the Faddeev-Popov charge -1 sector ((I) and \cite{bp2}). We have previously
seen that this relies on the existence of Killing vectors for the metric
$g_{ij}[\Phi]$ ; this is natural as such vectors signal extra isometries that
constrain the invariant action or, equivalently, signal the non physically
relevant character of some of the parameters of the classical action that may
be reabsorbed through a conveniently chosen field and source reparametrisation
\cite{7}.

As in the previous section, the analysis separates with respect to the number
$N_{d_{\al}}$ :

\subsubsection{$N_{d_{\al}} = 0$}
The image of $\De_{[0]}^0$ (\ref{a81}) through $S_L^r$ does not intercept
$\De^{an.(0)}_{[+1]}$, the cohomology of $S_L^{00}$ in the anomaly sector. As a
consequence (\cite{14} and the appendix of (I)), there will be no obstruction
in the construction of the cocycles of $S_L^0$ starting from those of
$S_L^{00}$ and there is an isomorphism between the two cohomology spaces.
Consequently, the cohomology in the $N_{d_{\al}} =0$ Faddeev-Popov neutral
sector is characterized by a symmetric tensor $t_{ij}[\Phi]$ such that $g'_{ij}
= g_{ij} + \hbar t_{ij}$ is a metric, hermitian with respect to the very
complex structure $J^i_j$ we started from, and such that $J^i_j$ is covariantly
constant with respect to the covariant derivative with connexion
$\Ga^k_{ij}[g'_{mn}]$. This is the necessary stability of the N=2
supersymmetric theory which ensures that, at a given perturbative order where
the Slavnov identity holds (absence of anomaly up to this order), the U.V.
divergences in the Green functions may be compensated for through the usual
renormalisation algorithm and normalisation conditions \cite{6}.

\subsubsection{$N_{d_{\al}} =1$}
Here the image of $\De_{[0]}^1$ (\ref{a81}) through $S_L^r$ intercepts
$\De^{an.(1)}_{[+1]}$, the cohomology of $S_L^{00}$ in the anomaly sector. As a
consequence (\cite{14} and the appendix of (I)), this will restrict the
cohomology (\ref{a812},\ref{a91}) in the considered sector. In fact, $S_L^{0}
\De_{[0]}^1 = 0 $ gives (no coboundaries exist in that sector) :
\beqa\label{G1}
\De_{[0]}^1 & = & \De_{[0]}^{an.(1)}[U^i_{\al\,j}(\Phi),\
S_{\al}^{ij}(\Phi)]_{[equ. (\ref{a81})]}\ \nnb\\
 {\rm with}\ \ U^i_{\al\,j}(\Phi) \ = \ -2[J_{jk} S_{\al}^{ki}(\Phi)] &
\Leftrightarrow & S_{\al}^{ij} = -{1\over 2}J^{ik}U^j_{\al\,k}
\eeqa
where the supplementary constraint on  $S_{\al}^{ij}$ is such that
$J^i_{\al\,j} + \hbar U^i_{\al\,j}$ - which anticomutes with $J^i_{j}$ - is now
also covariantly constant with respect to the covariant derivative with
connexion $\Ga^k_{ij}[g_{mn}]$.

\subsubsection{$N_{d_{\al}} =2$}
Here too, the image of $\De_{[0]}^2$ (\ref{a81}) through $S_L^r$   intercepts
$\De^{an.(2)}_{[+1]}$, the cohomology of $S_L^{00}$ in the anomaly sector,
which will restrict the cohomology (\ref{a813},\ref{a92}) in the considered
sector. Thanks to the simplicity of $\De_{[0]}^2$ (\ref{a81}), the analysis of
the cocycle condition $S_L^0\De_{[0]}^2 = 0$ in the one $\eta_i$ subsector
readily shows that the cohomology space of the complete $S_L^0$ operator is
empty in this $N_{d_{\al}} =2$, Faddeev-Popov neutral sector.

Finally, let us consider the Faddeev-Popov charge +1 sector.

\subsection{The Faddeev-Popov +1 charge sector}

Here too, the analysis  separates with respect to the number $N_{d_{\al}}$.

\subsubsection{$N_{d_{\al}} =0$}

The $S_L^{00}$ cohomology was obtained in equ.(\ref{b91}) :
$$\De_{[+1]}^{an.(0)} = \int d^2x d^2\th t^{[ijk]}[\Phi](d^+)^2(d^-)^2
\eta_i\eta_j\eta_k $$

\noindent and, using the algorithm described by equ.(\ref{a551}), we find the
$S_L^0$ cohomology in the same sector to be :
\beqa\label{a12}
\De_{[+1]}^{an.(0)} &=& \int d^2x d^2\th t^{[ijk]}[\Phi]
\{(d^+)^2(d^-)^2\eta_i\eta_j\eta_k - {3\over 2}d^+d^-\eta_i\eta_j
J_{kn}(d^+D_+\Phi^n - d^-D_-\Phi^n)\nnb\\
&-& 3d^+d^-\eta_i J_{jn} J_{km}D_+\Phi^nD_-\Phi^m \nnb\\
&+& {3\over 4}  J_{in} J_{jm} J_{kl}(d^+D_+\Phi^n D_+\Phi^m D_-\Phi^l -
d^-D_-\Phi^n D_-\Phi^m D_+\Phi^l) \} \nnb\\
&+& \int d^2x d^2\th t^1_{[nm]l}[\Phi]\left(d^+D_+\Phi^n D_+\Phi^m D_-\Phi^l -
d^-D_-\Phi^n D_-\Phi^m D_+\Phi^l\right) \ ,
\eeqa
where $t^1_{[ij]\,k}[\Phi]$ is related to the pure contravariant analytic
tensor $t^{[ijk]}[\Phi]$ through (in complex coordinates) :
$$t^1_{[ab]\,\bar{c}},\ \ t^1_{[\bar{a}\bar{b}]\,c} \neq 0,\ \ {\rm \ the\
other\ vanish }\ ;$$
$$t^1_{[ab]\,\bar{c}} = {i\over 4} \partial_{\bar{c}}[g_{a\bar{a}}g_{b\bar{b}}
K,_{\bar{d}}t^{[\bar{a}\bar{b}\bar{d}]}] \ \ {\rm where\ K\ is\ the\
K\ddot{a}hler\ potential}\ .$$

\subsubsection{$N_{d_{\al}} =1$}

The $S_L^{00}$ cohomology was obtained in equ.(\ref{b911}) :
$$\De_{[+1]}^{an.(1)} = \int d^2x d^2\th t_{\al}^{[ijk]}(\Phi)
[d_{\al}^+d^+(d^-)^2 + d_{\al}^-d^-(d^+)^2]\eta_i\eta_j\eta_k \ +$$
$$+  \int d^2x d^2\th \tilde{t}_{\al\,[ij]k}(\Phi)
[d_{\al}^+D_+\Phi^iD_+\Phi^jD_-\Phi^k - d_{\al}^-D_-\Phi^iD_-\Phi^jD_+\Phi^k
]$$

\noindent Notice first that $S_L^r$ does not act on the second piece of
$\De_{[+1]}^{an.(1)}$ , which then is a true $S_L^0$ anomaly. Due to its
similarity with $\De_{[+1]}^{an.(0)}$, the first part of $\De_{[+1]}^{an.(1)}$
is easily promoted to a complete $S_L^0$ cohomology :
\beqa\label{a121}
\De_{[+1]}^{an.(1)} &=& \int d^2x d^2\th t_{\al}^{[ijk]}(\Phi)
\{d_{\al}^+d^+(d^-)^2\eta_i\eta_j\eta_k - {3\over 2}d_{\al}^+d^-\eta_i\eta_j
J_{kn}(d^+D_+\Phi^n - d^-D_-\Phi^n)\nnb\\
&-& 3d_{\al}^+d^-\eta_i J_{jn} J_{km}D_+\Phi^nD_-\Phi^m + {3\over 2} J_{in}
J_{jm} J_{kl}d_{\al}^+D_+\Phi^n D_+\Phi^m D_-\Phi^l +  {\rm parity\ exchange }
\} \nnb\\
&+& \int d^2x d^2\th (2t^1_{\al[ij]k}(\Phi) + \tilde{t}_{\al[ij]k}(\Phi))
\left(d_{\al}^+D_+\Phi^i D_+\Phi^j D_-\Phi^k  -
d_{\al}^-D_-\Phi^iD_-\Phi^jD_+\Phi^k \right)
\eeqa
where $t^1_{\al[ij]\,k}[\Phi]$ is related to the pure contravariant analytic
tensor $t_{\al}^{[ijk]}[\Phi]$ as in the $N_{d_{\al}} =0$ case, and
$\tilde{t}_{\al[ij]k}[\Phi]$ is related to a covariant vector $t_{\al\,i}$ (see
subsection 3.3.2).

\subsubsection{$N_{d_{\al}} =2$}

The $S_L^{00}$ cohomology was obtained in equ.(\ref{b912}):
$$\De_{[+1]}^{an.(2)} = \De_{[+1]}^{an.(2)}|_{(\al\be)++} +
\De_{[+1]}^{an.(2)}|_{(\al\be)--} + \De_{[+1]}^{an.(2)}|_{(\al\be)+-} +
\De_{[+1]}^{an.(2)}|_{[\al\be]} $$
where :
$$ \De_{[+1]}^{an.(2)}|_{(\al\be)++} = d_{\al}^+d_{\be}^+ \int d^2x d^2\th
\eta_i\left[(d^-)^2\eta_j\eta_kt_{(\al\be)}^{[ijk]}(\Phi) +
(D_+)^2\Phi^jt_{(\al\be)\, j}^i(\Phi)\right] $$

$$ \De_{[+1]}^{an.(2)}|_{(\al\be)+-} = d_{\al}^+d_{\be}^-d^+d^- \int d^2x
d^2\th \eta_i\eta_j\eta_k\tilde{t}_{(\al\be)}^{[ijk]}(\Phi) $$

$$\De_{[+1]}^{an.(2)}|_{[\al\be]} =  d_{\al}^+d_{\be}^- \int d^2x d^2\th
\eta_i\left[D_+\Phi^jD_-\Phi^k t_{[\al\be]\,(jk)}^{i}(\Phi) + D_+D_-\Phi^j
t_{[\al\be]\,j}^{i}(\Phi)\right]$$

$\bullet$ As $S_L^{00}\De_{[+1]}^2$ contains at least one source $\eta_i$, and
$S_L^r$ decreases the number of $\eta_i$ of one unity, the $[\al,\be]$
skew-symmetric part of the looked-for cocycles of $S_L^0$,
$\De_{[+1]}^2|_{[\al\be]}$ should satisfy
$$\left(S_L^r\De_{[+1]}^2\right)|_{no\ \eta} = 0$$
which readily gives :
$$ \left(\De_{[+1]}^2|_{[\al\be]}\right)_{1\eta} = S_L^r
\left(d_{\al}^+d_{\be}^- \int d^2x d^2\th
\eta_i\eta_jT_{[\al\be]}^{[ij]}\right)\ .$$
The cocycle condition $(S_L^{00} + S_L^r)\De_{[+1]}^{an.(2)}|_{[\al\be]} = 0$
constrains $T_{[\al\be]}^{[ij]}$ to be a pure contravariant analytic
skew-symmetric tensor ; then, as  a consequence of (\ref{a813},\ref{a92}), the
last parenthesis is anihilated by  $S_L^{00}$ and $ \De_{[+1]}^2|_{[\al\be]}$
is in fact a $S_L^0$ coboundary.

$\bullet$ In the same way, the cocycle condition
$S_L^0\De_{[+1]}^2|_{(\al\be)+-} = 0$, when analysed by increasing number of
sources $\eta$, leads to $\De_{[+1]}^{an.(2)}|_{(\al\be)+-} = 0\ .$

$\bullet$ Finally, the cocycle condition  $S_L^0\De_{[+1]}^2|_{(\al\be)++} = 0$
is analysed along the same lines as $S_L^0\De_{[+1]}^1 = 0$, and leads to :
\beqa\label{a122}
\De_{[+1]}^{an.(2)} &=& d_{\al}^+d_{\be}^+\int d^2x d^2\th
t_{(\al\be)}^{[ijk]}[\Phi] \left\{(d^-)^2\eta_i\eta_j\eta_k + {3\over
2}[\eta_i\eta_j J_{kn} d^-D_+\Phi^n + \eta_i J_{jn} J_{km}D_+\Phi^nD_+\Phi^m
]\right\}\nnb\\
&+& d_{\al}^+d_{\be}^+\int d^2x d^2\th \eta_i
t_{(\al\be)j}^i[\Phi](D_+)^2\Phi^j \nnb\\
& + & {\rm parity\ exchange\ ,}
\eeqa
where :

- as usual, $t_{(\al\be)}^{[ijk]}[\Phi]$ is a pure contravariant analytic
skew-symmetric tensor, further constrained so as the corresponding
$t^1_{(\al\be)[ij]k}$ tensor actually vanishes, $i.e.$ in complex coordinates,
$$t_{(\al\be)}^{[abc]} = t_{(\al\be)}^{[abc]}(\phi^d)\ \ ; \ \
t_{(\al\be)\,[abc]} \stackrel{\rm def.}{=} g_{a\bar{a}}g_{b\bar{b}}g_{c\bar{c}}
t_{(\al\be)}^{[\bar{a}\bar{b}\bar{c}]}(\bar{\phi}^{\bar{d}}) =
t_{(\al\be)\,[abc]}(\phi^d)$$
and the complex conjugate relations ;

- $t_{(\al\be)j}^i[\Phi]$ is a mixed analytic tensor, further constrained so as
the tensor $G_{(\al\be)\,ij} = g_{ik}t_{(\al\be)j}^k$ is a symmetric,
covariantly constant tensor (hermitian with respect to the complex structure
$J^i_j$ due to previous relations (subsection 3.3.3)). This will be important
in the following (subsection 5.1.6).
\subsubsection{$N_{d_{\al}} = 3$}

There are no coboundaries in that sector, and the $S_L^{00}$ cohomology was
obtained in equ.(\ref{b913}).
Notice that $S_L^{00}\De_{[+1]}^3$ contains at least two sources $\eta_i$ and
that $S_L^r$ decreases the number of $\eta_i$ of one unity ; then, when
analysed by increasing number of sources $\eta$,  the cocycle condition
$S_L^0\De_{[+1]}^3 = 0$, leads to $\De_{[+1]}^3 = 0\ .$

\subsubsection{$N_{d_{\al}} = 4$}

There are no coboundaries in that sector, and the $S_L^{00}$ cohomology was
obtained in equ.(\ref{b914}).
Notice that $S_L^{00}\De_{[+1]}^4$ contains at least three sources $\eta_i$ and
$S_L^r$ decreases the number of $\eta_i$ of one unity ; here again, when
analysed by increasing number of sources $\eta$,  the cocycle condition
$S_L^0\De_{[+1]}^4 = 0$, leads to $\De_{[+1]}^4 = 0\ .$
$$ $$
To sum up, the cohomology space of $S_L^0$ in the Faddeev-Popov charge +1
sector depends on skew-symmetric contravariant analytic 3-tensors
$t^{[ijk]}[\Phi],\ t_{\al}^{[ijk]}[\Phi]$ and $t_{(\al\be)}^{[ijk]}[\Phi]$
, the last one endowing  a further constraint, on a vector $t_{\al\,i}$ and on
a symmetric, covariantly constant tensor $G_{(\al\be)\,ij}$, hermitian with
respect to the complex structure $J^i_j\ .$

We are now in a position to compute the cohomology of the complete $S_L = S_L^0
+ S_L^1 + S_L^2$ operator. But, as an intermediate result, we comment on N = 2
supersymmetric non-linear $\si$ models \cite{bo2}.

\subsection{N=2 supersymmetric non-linear $\si$ models}

In the special case of N=2 supersymmetric non-linear $\si$ models, there is no
need for ghosts $d_{\al}^{\pm}$ and all the necessary results may be found in
subsections 4.1, 4.2.1 and 4.3.1. In particular, equation (\ref{a12}) offers a
candidate for an anomaly and, as a consequence, if at a given pertubative order
this anomaly appears with a non zero coefficient
$$S_L^0\Ga|_{p^{th} order} = a (\hbar)^p \De_{[+1]}^{an.}, \ \ a \neq 0 $$
the N = 2 supersymmetry is broken as $\De_{[+1]}^{an.}$ cannot be reabsorbed
(being a  cohomology element, it is not a $S_L^0 \tilde{\De}_{[0]}$) and, {\it
a priori}, we are no longer able to analyse the structure of the U.V.
divergences at the next perturbative order, which is the death of the theory.

We now discuss some properties of the candidate anomaly, trying to characterize
the special geometries (and manifolds $\cal{M}$) where such pure contravariant
analytic skew-symmetric 3-tensor cannot appear.

Consider the covariant tensor
$$t_{[abc]} =
g_{a\bar{a}}g_{b\bar{b}}g_{c\bar{c}}t^{[\bar{a}\bar{b}\bar{c}]}[\bar{\phi}]\
.$$
It satisfies $\nabla _{d}\,t_{[abc]} = 0\ .$ The (3-0) form
\beq\label{A2}
\om ' = \frac{1}{3!}t_{[abc]}d\phi^{a}\wedge d\phi^{b} \wedge d\phi^{c}
\eeq
which satisfies $d'\om ' = 0$ \footnote{\ As usual in complex geometry, see for
example (\cite{W16},\cite{16}), the differential d $\equiv$ d' + d", the
codifferential $\de \equiv \de '+\de " $ and the Laplacian $\triangle \equiv
d\de +\de d$ .}, will now be shown to be harmonic if $\cal{M}$ is a
\underline{compact} manifold (or if it is a \underline{Ricci-flat} one - for
example an HyperK\"ahler manifold).

One firstly obtains from the identity :
$$\triangle t_{ijk} = g^{mn} \nabla _{m}\nabla _{n}\,t_{[ijk]}
-[R^{l}_{i}\,t_{[ljk]} + \ {\rm perms.}\ ] - [R^{lm}_{ij}\,t_{[lmk]} + \ {\rm
perms.}\ ]$$
rewriten in complex coordinates and due to $\nabla _{d}\,t_{[abc]} = 0$ :
$$\triangle t_{[abc]} = g^{d\bar{d}} \nabla _{d}\nabla _{\bar{d}}\,t_{[abc]}
-[R^{d}_{a}\,t_{[dbc]} + \ {\rm perms.}\ ]\ .$$
On another hand, the Ricci identity gives, still using $\nabla _{d}\,t_{[abc]}
= 0$ :
$$g^{d\bar{d}} \nabla _{d}\nabla _{\bar{d}}\,t_{[abc]} =
g^{d\bar{d}}[R^{e}_{\bar{d}da}\,t_{[ebc]} + \ {\rm perms.}\ ] =
-[R^{d}_{a}\,t_{[dbc]} + \ {\rm perms.}\ ].$$
Then $ \triangle t_{[abc]} = 2g^{d\bar{d}} \nabla _{d}\nabla
_{\bar{d}}\,t_{[abc]}\ = \-2[R^{d}_{a}\,t_{[dbc]} + \ {\rm perms.}\ ]\ $. In
the Ricci-flat case, this gives the claimed harmonicity of $\om '$. Now, when
the manifold is a compact one, one may compute :
\beqa\label{A1}
(d\om ',d\om ') + (\de\om ',\de\om ') = (\om ',(d\de+\de d)\om ') = (\om
',\triangle\om ') & = & \int_{\cal{M}} d\si 2t^{[abc]}g^{d\bar{d}} \nabla
_{d}\nabla _{\bar{d}}\,t_{[abc]} =\nnb\\
= \int_{\cal{M}} d\si 2g^{d\bar{d}} \{\nabla _{d}\nabla
_{\bar{d}}(t^{[abc]}t_{[abc]}) - \nabla _{d}\,t^{[abc]}\nabla
_{\bar{d}}\,t_{[abc]}\}  & = & 0 - 2(d\om ',d\om ')\nnb\\
\Rightarrow \ (\de\om ',\de\om ') + 3(d\om ',d\om ') = 0
\ & \Rightarrow &\ \de\om ' = d\om '= \triangle \om' = 0\ ,
\eeqa
and, as a consequence :
\beq\label{A11}
[R^{d}_{a}\,t_{[dbc]} + \ {\rm perms.}\ ] = 0 \ \Rightarrow \
3t^{[abc]}R^{a'}_a\,t_{a'bc} = 0\ .
\eeq
Moreover, in this compact case, it results from the previous discussion that
$\left[\partial_{d}t^1_{[ab]\,\bar{c}} \right]_{(a,b,d\ a.s.)} \equiv {i\over
4}\partial_{\bar{c}}t_{[abd]}$ vanishes. As a consequence, $t^1_{[ab]\,\bar{c}}
= \partial_{b}\bar{t}_{a\bar{c}} - (a \leftrightarrow b)$, which corresponds to
a trivial cohomology and is then thrown away from equation (\ref{a12}).

\noindent Notice that, as $a\ priori\ d\om '\neq 0$, this harmonicity is only a
necessary condition which, to my knowledge, is not given in the mathematics
litterature. For example, in \cite{W16}, page 70, only a more restrictive
necessary and sufficient condition is given :

{\sl On a compact K\"ahlerian manifold, given a (p,0) form $\eta$, the 3
following  conditions are equivalent : $\eta$ is closed, $\eta$ is holomorphic
and $\eta$ is harmonic,}
\newline or in \cite{16}, theorem 9.3 :

{\sl On a compact K\"ahlerian manifold, given a pure skew-symmetric covariant
tensor, a necessary and sufficient condition for it to be analytic is that it
be harmonic.}
\newline Other necessary and sufficient conditions depend on the sign of the
Ricci tensor : for example \cite{16}, theorem 9.6 :

{\sl If the Ricci tensor is positive definite, there exists no contravariant
analytic tensor.} This results immediatly from (\ref{A11}).

\noindent It is known that the number of such harmonic (3,0) forms is given by
the Hodge number $h^{(3,0)}$; then this number determines an upper bound for
the dimension of the cohomology space of $S_L^0$ in the anomaly sector.

As a first result, this proves that if the manifold $\cal{M}$ has a complex
dimension smaller than 3, there is no anomaly candidate.

\noindent Another special case is the compact K\"ahler homogeneous one (N=2
supersymmetric extension of our previous works (I) for N=1 susy and \cite{7}
for the purely bosonic case) : in such a case the Ricci tensor is positive
definite \cite{77} which, due to the aforementionned theorem, forbids the
existence of such analytic tensor $t^{[abc]}(\phi^{d})$. As a consequence, the
cohomology of $S_L^{00}$ - and then of $S_L^0$ - vanishes in the anomaly
sector, $i.e.$ the Slavnov identity is not anomalous, which means that, as
expected, N=2 supersymmetry is renormalisable (at least in the absence of
torsion).

Moreover, due to the ``stability" \footnote{\ The trivial cohomology
$S_L^0\De_{[-1]}$ corresponds to field and source reparametrisations according
to (\ref{C7},\ref{C71}).} of the complex structure and of the classical action
in the space of K\"ahler metrics (subsection 4.2.1), in this case of d=2, N=2
non-linear $\si$ models, the renormalisation algorithm {\sl a priori} does not
change the number of parameters with respect to the one of the classical
action. Of course, as mentionned in subsections 4.1 and 4.2, in the presence of
Killing vectors, $i.e.$ of extra isometries, the generic symmetric, K\"ahlerian
metric tensor $t_{ij}[\Phi]$ gets some more constraints. For example, when the
manifold is an homogeneous K\"ahler one (usual non linear $\si$ models on coset
spaces), up to infra-red analysis, our work extends to the N=2 supersymmetric
case the renormalizability proof given for the bosonic case in \cite{7} (and
for the N=1 case in Section 5 of (I)).

Of course, when $h^{(3,0)} \neq 0$ - which for example occurs for Calabi-Yau
manifolds, $i.e.$ compact Ricci-flat K\"ahler manifolds of complex dimension 3,
where $h^{(3,0)} = 1$(ref.\cite{2})\footnote{\ As $\det\|g\| = 1$, a
representative of $t^{[abc]}$ is the constant skew-symmetric tensor
$\epsilon^{[abc]}$( with $\epsilon^{123} = +1$).}  -, we have a true anomaly
candidate \cite{bo2}. Of course, as no explicit metric is at hand, one cannot
compute the anomaly coefficient.

\noindent Some comments on its possible vanishing in perturbation theory will
be offered in the Concluding Section.

This anomaly in \underline{global} extended supersymmetry is a surprise with
respect to common wisdom \cite{18} (but see other unexpected non-trivial
cohomologies in supersymmetric theories in the recent works of Brandt \cite{8}
and Dixon \cite{9}) and the fact that if we have chosen, from the very
beginning, a coordinate system adapted to the complex structure, the second
supersymmetry will be linear and there will be no need for sources $\eta_i$ .
However, as known from chiral symmetry, even a linearly realised transformation
can lead to anomalies ; moreover, here the linear supersymmetry transformations
do not correspond to an ordinary group but rather to a supergroup where,
contrarily to ordinary compact groups \footnote{\ In the appendix A of ref.
\cite{7}, it is proven that any linearly realised symmetry corresponding to a
\underline{compact} group can be implemented to all orders of perturbation
theory.} no general theorems exists : then there is no obvious contradiction.
This emphasizes the special structure of the supersymmetry algebra.

\section {The cohomology of $S_L$ and N=4 supersymmetry}
In subsection 2.1, we have split the complete linearised B.R.S. operator $S_L$
into 3 pieces, according to their number of ghosts $d^{\pm}_{\al}$ :
$$S_L = S_L^0 + S_L^1 + S_L^2$$
where :
\beqa\label{h1}
S_L^0 &=&  \int d^2x d^2\th\{J^i_j (d^+ D_+\Phi^j + d^- D_-\Phi^j)
\frac{\de}{\de\Phi^i} \nnb\\
&+& [ \frac{\de A^{inv.}}{\de \Phi^i} +\eta_k (J^k_{j\,,i}-J^k_{i\,,j})(d^+
D_+\Phi^j + d^- D_-\Phi^j) + J^j_i(d^+ D_+\eta_j + d^- D_-\eta_j)]
\frac{\de}{\de\eta_i}\}\ ,\nnb\\
S_L^1 &=&  \int d^2x d^2\th\{\left[J^i_{\al\,j}(d_{\al}^+ D_+\Phi^j + d_{\al}^-
D_-\Phi^j) + \eta_j\epsilon_{\al\be 3}J^{ij}_{\be}(d_{\al}^+d^- -
d_{\al}^-d^+)\right] \frac{\de}{\de\Phi^i} \nnb\\
&+& [J^j_{\al\,i}(d_{\al}^+ D_+\eta_j + d_{\al}^- D_-\eta_j) +\eta_k
(J^k_{\al\,j\,,i} - J^k_{{\al\,i\,,j}})(d_{\al}^+ D_+\Phi^j + d^- D_-\Phi^j)
+\nnb \\
& + & {1\over 2}\eta_k\eta_l\epsilon_{\al\be 3}J^{kl}_{\be\,,i} (d_{\al}^+d^- -
d_{\al}^-d^+)] \frac{\de}{\de\eta_i}\}\ ,\nnb\\
{\rm and}\ S_L^2 &=&  -\epsilon_{\al\be 3}d_{\al}^+d_{\be}^-\int d^2x d^2\th
\left[\eta_jJ^{ij}\frac{\de}{\de\Phi^i} + {1\over 2}\eta_k\eta_lJ^{kl}_{\ ,i}
\frac{\de}{\de\eta_i}\right]\ ,
\eeqa
and the cohomology of $S_L^0$ was obtained in the previous Section. We start
the analysis by the anomaly sector, as the existence of a true Slavnov anomaly
would be the death of the theory.

\subsection{ The Faddeev-Popov +1 charge sector}

Filtering the cocycle condition $S_L\De_{[+1]} = 0$ with respect to the number
$N_{d_{\al}^{\pm}}$, we know that the cohomology of $S_L$ will be a subspace of
the one of $S_L^0$. We now analyse this restriction of the cohomology space
(due to a possible Faddeev-Popov +2 charge anomaly) in successive filtration
orders.

\subsubsection{ $N_{d_{\al}} = 0$}
The cocycle condition  $S_L^0\De^0_{[+1]} = 0$ has been solved in subsection
4.3.1 (equ.(\ref{a12})) :
$$\De^0_{[+1]} = \De_{[+1]}^{an.(0)}[t^{[ijk]}(\Phi)] + S_L^0\De^0_{[0]} $$

\subsubsection{$N_{d_{\al}} =1$}
At this order, and using previous result, the cocycle condition writes :
\beq\label{h10}
S_L^0(\De^1_{[+1]} - S_L^1\De^0_{[0]}) = -
S_L^1\De_{[+1]}^{an.(0)}[t^{[ijk]}(\Phi)]
\eeq
{}From $S_L^0(S_L^1\De_{[+1]}^{an.(0)}) = - S_L^1(S_L^0\De_{[+1]}^{an.(0)}) =
0$, one sees that (\ref{h10}) $a\ priori$ enforces some constraints on the
tensor $t^{[ijk]}(\Phi)\ .$ From the condition :
\beq\label{h11}
S_L^1\De_{[+1]}^{an.(0)}[t^{[ijk]}(\Phi)] = S_L^0\tilde{\De}_{[+1]}^1
\eeq
where the general expression for $ \tilde{\De}_{[+1]}^1$ was given in
(\ref{b101}), one obtains at decreasing orders in the number of ghosts, firstly
a constraint \footnote{\ In the \underline{compact} K\"ahler case, we have
shown in section 4.4 that $\nabla_l t^{[ijk]} = 0$ (as a consequence of $d\om'
= 0$) . Then this is not a new constraint.} on $t^{[ijk]} $ :
$$N_{d_{A}} = 6\ \ :\ \ \nabla_i [J_{\al}^{ij}t^{[klm]}]|_{[j,k,l,m]\,a.s.} =
0\ ,$$
then $t_{\al}^{[ijk]}(\Phi)$ in function of $t^{[ijk]}(\Phi)$. Using complex
coordinates - in particular the fact that
$J^d_{\al\,\bar{d}}\,(\phi,\bar{\phi}) =
\partial_{\bar{d}}J^d_{\al}\,(\phi,\bar{\phi})$ - , this writes :
$$\partial_{\bar{d}}t_{\al}^{[abc]} = {i\over 4}J^d_{\al\,\bar{d}}\nabla_d
t^{[abc]}(\phi) = \partial_{\bar{d}}\left[{i\over
4}J^d_{\al}(\phi,\bar{\phi})\nabla_d t^{[abc]}(\phi)\right]\ ,$$
$$t_{\al}^{[\bar{a}bc]} = {i\over
4}J^{\bar{a}}_{\al\,a}(\phi,\bar{\phi})t^{[abc]}(\phi)\ ,$$
which exhibits some arbitrariness in $\tilde{\De}_{[+1]}^1$, defined up to a
$S_L^0$ cocycle $\De_{[+1]}^{an.(1)}|_{equ. (\ref{a121})} + S_L^0\De^1_{[0]}\
.$ At the next order ($N_{d_{A}} = 4$) we obtain the vanishing of
$t^{[ijk]}(\Phi)\ i.e.$ of $\De_{[+1]}^{an.(0)}\ .$

As a consequence, equ.(\ref{h10}) writes :
$S_L^0(\De^1_{[+1]} - S_L^1\De^0_{[0]}) = 0 $ which, according to subsection
4.3.2 (equ.(\ref{a121})) solves to :
\beq\label{h12}
\De^1_{[+1]} =
\De_{[+1]}^{an.(1)}[t_{\al}^{[ijk]}(\Phi),\,\tilde{t}_{\al\,[ij]k}(\Phi)] +
S_L^0\De^1_{[0]} + S_L^1\De_{[0]}^0\ .
\eeq

\subsubsection{$N_{d_{\al}} =2$}
At this order, and using previous results, the cocycle condition writes :
\beq\label{h20}
S_L^0(\De^2_{[+1]} - S_L^2\De^0_{[0]} - S_L^1\De^1_{[0]}) = -
S_L^1\De_{[+1]}^{an.(1)}[t_{\al}^{[ijk]},\,\tilde{t}_{\al\,[ij]k}]
\eeq
Here too, from $S_L^0(S_L^1\De_{[+1]}^{an.(1)}) = -
S_L^1(S_L^0\De_{[+1]}^{an.(1)}) = 0$, one sees that (\ref{h20}) $a\ priori$
enforces some constraints on the tensors $t_{\al}^{[ijk]}$ and
$\tilde{t}_{\al\,[ij]k}.$ From the condition :
\beq\label{h21}
S_L^1\De_{[+1]}^{an.(1)}[t_{\al}^{[ijk]},\,\tilde{t}_{\al\,[ij]k}] =
S_L^0\tilde{\De}_{[+1]}^2
\eeq
where the general expression for $ \tilde{\De}_{[+1]}^2$ was given in
(\ref{b102}) and is defined up to the cocycles
$\De_{[+1]}^{an.(2)}|_{equ.(\ref{a122})} + S_L^0\De^2_{[0]}\ ,$ the same
analysis as in the previous subsection for $N_{d_{A}}$ = 6, 5 and 4 again gives
the vanishing of $t_{\al}^{[ijk]}(\Phi)$ and the ``triviality" of the
corresponding terms of $\tilde{\De}_{[+1]}^2$ , as they reduce themselves to
$S_L^0$ cocycles. Then  $\De_{[+1]}^{an.(1)}$ depends only on the tensor $
\tilde{t}_{\al\,[ij]k}(\Phi)$. So $S_L^1\De_{[+1]}^{an.(1)}$ involves at most 3
ghosts and in $\tilde{\De}_{[+1]}^2$, only the terms in
$t^i_{(\al\be)\,4[jk]}$, $t'^i_{(\al\be)\,4[jk]}$, $t^i_{[\al\be]\,4(jk)}$ and
$t^i_{(\al\be)\,5j}$, $t^i_{[\al\be]\,5j}$ survive.

When analysed at ghost level 3 and 2, the condition (\ref{h21}) does not lead
to the vanishing of these tensors, but expresses them as functions of the
complex structure $J^i_{\al\,j}$ and the ``anomaly tensor"
$\tilde{t}_{\al\,[ij]k}.$ For example, one finds (in complex coordinates) :
$$t^a_{(\al\be)\,4[bc]} = t^a_{(\al\be)\,4[b\bar{c}]} = 0\ ,\
t^a_{(\al\be)\,4[\bar{b}\bar{c}]} = {1\over
4}\left(J^{ad}_{\al}\tilde{t}_{\be\,[\bar{b}\bar{c}]d} + (\al \leftrightarrow
\be) \right)\ ,$$
$$t^a_{(\al\be)\,5\,\bar{b}} = 0\ ,\  t^a_{(\al\be)\,5\,b\,,\bar{c}} = {1\over
2}\left(J^{ad}_{\al}\tilde{t}_{\be\,[db]\bar{c}} + (\al \leftrightarrow \be)
\right)\ ,$$
and constraints such as  :
$$\left( J^{\bar{a}}_{\al\,b}\tilde{t}_{\be\,[cd]\bar{a}} +
J^{\bar{a}}_{\al\,c}\tilde{t}_{\be\,[db]\bar{a}} +
J^{\bar{a}}_{\al\,d}\tilde{t}_{\be\,[bc]\bar{a}} + (\al \leftrightarrow \be)
\right) = 0\ .$$
Finally, from (\ref{h20},\ref{h21}) and subsection 4.3.3, one gets :
\beq\label{h22}
\De^2_{[+1]} = \De_{[+1]}^{an.(2)}[t_{(\al\be)}^{[ijk]}, t^{i}_{(\al\be)\,5j}]
-\tilde{\De}^2_{[+1]} + S_L^0\De^2_{[0]} + S_L^1\De^1_{[0]} + S_L^2\De_{[0]}^0\
,
\eeq
where the tensor $G_{(\al\be)\,ij} = g_{ik}t_{(\al\be)\,5j}^k$ is a symmetric,
covariantly constant tensor, hermitian with respect to the complex structure
$J^i_j$ (subsection 4.3.3).

\subsubsection{$N_{d_{\al}} = 3$}

At this order, and using previous results, the cocycle condition writes :
\beqa\label{h30}
& S_L^0 &\left(\De^3_{[+1]} - S_L^2\De^1_{[0]} - S_L^1\De^2_{[0]}\right) =
-S_L^2\left(\De_{[+1]}^{an.(1)}[\tilde{t}_{\al\,[ij]k}]\right) -\nnb\\
 - & S_L^1 &
\left(\tilde{\tilde{\De}}_{[+1]}^2
[t_{(\al\be)}^{an.[ijk]},t^{an.\,i}_{(\al\be)\,5j}\ ;\ t^i_{(\al\be)\,4[jk]},
t'^i_{(\al\be)\,4[jk]},t^i_{[\al\be]\,4(jk)},
t^i_{(\al\be)\,5j},t^i_{[\al\be]\,5j}]\right)
\eeqa
where $\tilde{\tilde{\De}}_{[+1]}^2 = \De_{[+1]}^{an.(2)} -
\tilde{\De}_{[+1]}^2\ .$ Here too, from $S_L^0(S_L^2\De_{[+1]}^{an.(1)} +
S_L^1\tilde{\tilde{\De}}_{[+1]}^2) = - S_L^1 S_L^1\De_{[+1]}^{an.(1)} -
S_L^1S_L^0\tilde{\tilde{\De}}_{[+1]}^2 \stackrel{(\ref{h21})}{=} 0\,$, one sees
that (\ref{h30}) $a\ priori$ enforces some new constraints on the tensor
$\tilde{t}_{\al\,[ij]k}\ .$ From the condition :
\beq\label{h31}
S_L^2\De_{[+1]}^{an.(1)} + S_L^1\tilde{\tilde{\De}}_{[+1]}^2 =
S_L^0\tilde{\De}_{[+1]}^3
\eeq
where the general expression for $ \tilde{\De}_{[+1]}^3 $ was given in
(\ref{b103}), and thanks to the special structure of $S_L^2$ (\ref{h1}), one
obtains through analysis of the terms in $d^+_{\al}d^+_{\be}d^+_{\ga}$, at the
6 and 5 order in the total number of ghosts, the vanishing of the part in
$t_{(\al\be)}^{an.[ijk]}(\Phi)$ of the anomaly $\De_{[+1]}^{an.(2)}$ involved
in $
\tilde{\tilde{\De}}_{[+1]}^2\ .$

Then, the analysis of the terms in $d^+_{\al}d^+_{\be}d^+_{\ga}d^-$ gives :
\newline
$\bullet$ the symmetry, in the exchange i $\leftrightarrow$ j, of :
$\epsilon_{3\al\de}J_{\de}^{ki}\left[t^{an.\,j}_{(\be\ga)5\,k} +
t^j_{(\be\ga)5\,k}\right]|_{(\al\be\ga)\;sym.}\ ,$
\newline
$\bullet$ the vanishing of $t^k_{(\be\ga)4\,[ij]}\,,$ and then :
$\tilde{t}_{\al\,[ij]k}=0\ \ i.e.\ \ \De_{[+1]}^{an.(1)}=0\,.$

As a consequence, equ.(\ref{h21}) gives : $S_L^0\tilde{\De}_{[+1]}^2
 =0\,,$ and, thanks to previous remarks, $\tilde{\De}_{[+1]}^2$ may be
supressed, $i.e.\ t^i_{(\be\ga)\,5j} \equiv 0\,.$ Then, the constraint
(\ref{h31}) reduces itself to :
\beq\label{h32}
S_L^1\De_{[+1]}^{an.(2)}[t^{an.\,i}_{(\al\be)5\,j}(\Phi)] =
S_L^0\tilde{\De}_{[+1]}^3
\eeq
In the analysis of the terms in $d^-_{\al}d^+_{\be}d^+_{\ga}$, equ.(\ref{h32})
may be shown to enforce a stronger constraint on
$t^{an.\,i}_{(\al\be)5\,j}(\Phi)$ :
\beq\label{h33}
J_{\al\,k}^it^{an.\,k}_{(\be\ga)5\,j} = J_{\al\,j}^kt^{an.\,i}_{(\be\ga)5\,k}
\ \ \Leftrightarrow \ \ J_{\al\,k}^iG_{(\be\ga)\,ij} = -
G_{(\be\ga)\,ik}J_{\al\,j}^i\ \ \ \forall\ \al,\ \be,\ \ga\ ,
\eeq
$i.e.$ the tensor $G_{(\al\be)\,ij}$ is hermitian with respect to the complex
structures $J_{\al\,k}^i\,.$ Moreover, $\tilde{\De}_{[+1]}^3 \equiv 0\,.$
Finally, from (\ref{h30},\ref{h31}) and subsection 4.3.4, one gets :
\beq\label{h34}
S_L^1\De_{[+1]}^{an.(2)}=0\ \ ; \ \ \De^3_{[+1]} = S_L^1\De^2_{[0]} +
S_L^2\De^1_{[0]}\ .
\eeq

\subsubsection{$N_{d_{\al}} = 4$}

At this order, and using previous results, the cocycle condition writes :
\beq\label{h40}
S_L^0(\De^4_{[+1]} - S_L^2\De^2_{[0]}) = -
S_L^2\De_{[+1]}^{an.(2)}[t_{(\al\be)\,j}^{an.\,i}]
\eeq
Here too, from :
$$S_L^0(-S_L^2\De_{[+1]}^{an.(2)})  = S_L^1(S_L^1\De_{[+1]}^{an.(2)})
\stackrel{(\ref{h34})}{=} 0\ ,$$
one sees that (\ref{h40}) $a\ priori$ enforces some constraint on the tensor
$t_{(\al\be)5\,j}^{an.\,i}.$ From the condition :
\beq\label{h41}
S_L^2\De_{[+1]}^{an.(2)} = S_L^0\tilde{\De}_{[+1]}^4
\eeq
where $\tilde{\De}_{[+1]}^4$ was given in (\ref{b104}), and the structures of
$S_L^2$ (\ref{h1}) and $\De_{[+1]}^{an.(2)}$, one sees that the right hand side
is proportionnal to  $d^+_{\al}d^-_{\be}d^+_{\ga}d^-_{\de}$ when no such terms
may occur in the left hand side, which means that both sides should vanish.
As a consequence, $\tilde{\De}_{[+1]}^4$ is a $S_L^0$ cocycle, which means that
it vanishes (subsection 4.3.5).

Finally, from (\ref{h40},\ref{h41}) and subsection 4.3.5, one gets :
\beq\label{h44}
S_L^2\De_{[+1]}^{an.(2)} = 0\ \ ;\ \ \De^4_{[+1]} = S_L^2\De^2_{[0]}\ .
\eeq

\subsubsection{Absence of N=4 supersymmetry anomaly}

Putting the results of the previous subsections 5.1.1-5 altogether shows that
the cohomology space of $S_L$ in the Faddeev-Popov charge +1 sector depends on
a single symmetric tensor $G_{(\al\be)\,ij}[\Phi]$, hermitian with respect to
the 3 complex structures $ J^i_{A\,j}$ and covariantly constant :
\beqa\label{h60}
\De_{[+1]} & = & \De_{[+1]}^{an.(2)} + S_L\De_{[0]}\ \ {\rm where}\ \
S_L^0\De_{[+1]}^{an.(2)} = S_L^1\De_{[+1]}^{an.(2)} = S_L^2\De_{[+1]}^{an.(2)}
= 0\,,  \nnb\\
\De_{[+1]}^{an.(2)} & = & \int d^2x d^2\th \eta_ig^{ij}G_{(\al\be)\,jk}
\left(d_{\al}^+d_{\be}^+(D_+)^2\Phi^j + d_{\al}^-d_{\be}^-(D_-)^2\Phi^j\right)
\ .
\eeqa
This is reminiscent of the right hand side of the classical Slavnov identity
(\ref{C5}). As a matter of facts, it appears that $\De_{[+1]}^{an.(2)}$ is a
$S_L$ cocycle :
\beqa\label{h61}
\De_{[+1]}^{an.(2)} & = & S_L\left(\De_{[0]}[G_{(\al\be)\,ij}(\Phi)]\right)
\nnb\\
\De_{[0]} & = & \int d^2x d^2\th \{ a G_{(\al\al)\,ij}D_+\Phi^iD_-\Phi^j -
{1\over 2} \eta_i J_{\al}^{ik}G_{(\al\be)\,kj}(d_{\be}^+ D_+\Phi^j + d_{\be}^-
D_-\Phi^j) + \nnb\\
& + & {1\over 2} \eta_i\eta_j\epsilon_{ABC}d_{A}^+d_{B}^-[(a+{1\over
2})J^{ik}_{C} G_{(\ga\ga)\,kl}g^{lj}
- {1\over 2}J^{ik}_{\ga} G_{(\ga\, C)\,kl}g^{lj}]\}
\eeqa
where a is an arbitrary constant. Then there is no Faddeev-Popov +1 charge
cohomology, and we get the absence of supersymmetry anomaly in N=4 non-linear
$\si$ models in 2 space-time dimensions.

\subsection{The Faddeev-Popov neutral charge sector}

As explained before and in the appendix of (I), the existence of a non trivial
cohomology for $S_L^0$ in the Faddeev-Popov +1 charge sector, will restrict the
cocycles of $S_L$ in the  Faddeev-Popov 0 charge sector ; this is easily
understood as N=4 supersymmetry enforces new constraints on the arbitrary
K\"ahlerian metric $g_{ij}$ corresponding to N=2 supersymmetry.

We now filter the cocycle condition $S_L\De_{[0]}$ = 0 with respect to the
number $N_{d^{\pm}_{\al}}$ .

\subsubsection{$N_{d_{\al}} = 0$}
The cocycle condition  $S_L^0\De^0_{[0]} = 0$ has been solved in subsection
4.2.1 :
$$\De^0_{[0]} = \De_{0]}^{an.(0)}[t_{(ij)}(\Phi)] +
S_L^0\De^0_{[-1]}[V^i(\Phi)] $$

\subsubsection{$N_{d_{\al}} =1$}
At this order, and using previous result, the cocycle condition writes :
\beq\label{i10}
S_L^0(\De^1_{[0]} - S_L^1\De^0_{[-1]}[V^i]) = -
S_L^1\De_{[0]}^{an.(0)}[t_{(ij)}(\Phi)]
\eeq
{}From $S_L^0(S_L^1\De_{[0]}^{an.(0)}) = - S_L^1(S_L^0\De_{[0]}^{an.(0)}) = 0$,
and the existence of an $S_L^0$ anomaly $\De_{[+1]}^{an.\,(1)}\,,$ one sees
that (\ref{i10}) enforces some constraints on the tensor $t_{(ij)}(\Phi)\ :$
\beq\label{i11}
S_L^1\De_{[0]}^{an.(0)}[t_{(ij)}] =
S_L^0\tilde{\De}_{[0]}^1[\tilde{U}^i_{\al\,j},\tilde{S}_{\al}^{[ij]}]
\eeq
where the general expression for $ \tilde{\De}_{[0]}^1$ was given in
(\ref{a81}). As a consequence, equ.(\ref{i10}) writes :
$S_L^0(\De^1_{[0]} - S_L^1\De^0_{[-1]} + \tilde{\De}_{[0]}^1) = 0 $ which,
according to subsection 4.2.2 (see in particular equ.(\ref{G1})), solves to :
\beqa\label{i12}
\De^1_{[0]} & = & \De_{[0]}^{an.(1)}[U^{an.\,i}_{\al\,j},\;S_{\al}^{an.\,[ij]}]
- \tilde{\De}_{[0]}^1[\tilde{U}^i_{\al\,j},\tilde{S}_{\al}^{[ij]}] +
S_L^1\De_{[-1]}^0[V^i]\nnb\\
& \equiv &
\tilde{\tilde{\De}}_{[0]}^1 [\tilde{\tilde{U}}^i_{\al\,j},
\tilde{\tilde{S}}_{\al}^{[ij]}] + S_L^1\De_{[-1]}^0[V^i]\ .
\eeqa
$\tilde{\De}_{[0]}^1$ being defined by equation (\ref{i11}) up to a $S_L^0$
cocycle, $i.e.$ up to $\De_{[0]}^{an.(1)}$, the constraint (\ref{i11})
especially implies the \underline{hermiticity} of the perturbed metric $g'_{ij}
= g_{ij} + \hbar t_{(ij)}$ with respect to the perturbed complex structures
$J^{'i}_{\al\,j} = J^{i}_{\al\,j} + \hbar \tilde{\tilde{U}}^i_{\al\,j}$ and the
\underline{covariant constancy} of the later with respect to the covariant
derivative with connexion corresponding to $g'_{ij}\,.$ Moreover, the tensor
$\tilde{\tilde{S}}_{\al}^{[ij]}$ is a pure contravariant analytic
skew-symmetric tensor, with components given in complex coordinates by :
$$\tilde{\tilde{S}}_{\al}^{[ab]} = -{i\over
4}g^{a\bar{c}}\left[\tilde{\tilde{U}}^b_{\al\,\bar{c}} +
t_{\bar{c}d}J_{\al}^{db}\right] - (a \leftrightarrow b)\ \ \equiv
\tilde{\tilde{S}}_{\al}^{[ab]}(\phi^d)\ \ {\rm and\ the\ complex\ conjugate\
relation,}$$
(compare to equ.(\ref{G1})), and one obtains the relations :
$$T_{\al\,[ab]} = -{i\over
4}\left[g_{a\bar{c}}\tilde{\tilde{U}}^{\bar{c}}_{\al\,b} +
t_{a\bar{c}}J_{\al\,b}^{\bar{c}}\right] - (a \leftrightarrow b)\ \ \equiv
T_{\al\,[ab]}(\phi^d)$$
and
$$\tilde{\tilde{U}}^b_{\al\,\bar{a}} \left( {\rm resp.\ }
\tilde{\tilde{S}}_{\al}^{[ab]}\right) = -i\epsilon_{\al\be 3}
\tilde{\tilde{U}}^b_{\be\,\bar{a}} \left( {\rm resp.\ }
\tilde{\tilde{S}}_{\be}^{[ab]}\right)\,$$
(compare to equ.(\ref{C3}) which implies : $J^b_{\al\,\bar{a}} =
-i\epsilon_{\al\be 3}J^b_{\be\,\bar{a}}\,)$ and the complex conjugate
relations.

\subsubsection{$N_{d_{\al}} =2$}
At this order, and using previous results, the cocycle condition writes :
\beq\label{i20}
S_L^0(\De^2_{[0]} - S_L^2\De^0_{[-1]}) = - S_L^2\De_{[0]}^{an.(0)} -
S_L^1\tilde{\tilde{\De}}_{[0]}^1
\eeq
Here too, from
$$S_L^0\left(S_L^2\De_{[0]}^{an.(0)} + S_L^1\tilde{\tilde{\De}}_{[0]}^1\right)
= - S_L^1(S_L^1\De_{[0]}^{an.(0)} - S_L^0\tilde{\De}_{[0]}^1)
\stackrel{(\ref{i11})}{=} 0\,,$$
and the existence of an $S_L^0$ anomaly $\De_{[+1]}^{an.\,(2)}\,,$ one sees
that (\ref{i20}) enforces some constraints on the tensors $t_{(ij)},\
\tilde{\tilde{U}}^i_{\al\,j},\tilde{\tilde{S}}_{\al}^{[ij]}\ :$
\beq\label{i21}
S_L^2\De_{[0]}^{an.(0)}[t_{(ij)}] +
S_L^1\tilde{\tilde{\De}}_{[0]}^1 [\tilde{\tilde{U}}^i_{\al\,j},
\tilde{\tilde{S}}_{\al}^{[ij]}] =
S_L^0\tilde{\De}_{[0]}^2[\tilde{S}^{[ij]}_{[\al\be]}]
\eeq
where the general expression for $ \tilde{\De}_{[0]}^2$ was given in
(\ref{a81}). As a consequence of (\ref{i21}), equ.(\ref{i20}) writes :
$S_L^0(\De^2_{[0]} - S_L^2\De^0_{[-1]} + \tilde{\De}_{[0]}^2) = 0 $ which,
according to subsection 4.2.3 solves to :
\beq\label{i22}
\De^2_{[0]} = -\tilde{\De}_{[0]}^2[\tilde{S}^{[ij]}_{[\al\be]}] +
S_L^2\De_{[-1]}^0[V^i]\ .
\eeq
The constraint (\ref{i21}) especially implies the anticommutation of the
perturbed complex structures $J^{'i}_{\al\,j}$ and $J^{'i}_{\be\,j}$ and the
relation :
$$\tilde{S}^{[ij]}_{[\al\be]} = -{1\over 2}\epsilon_{\al\be
3}J_3^{im}t_{(mn)}g^{nj}\ .$$

\subsubsection{$N_{d_{\al}} = 3$}

At this order, and using previous results, the cocycle condition writes :
\beqa\label{i30}
S_L^2\De^1_{[0]} + S_L^1\De^2_{[0]} & = & 0 \nnb\\
\Leftrightarrow \ \   S_L^1\tilde{\De}_{[0]}^2[\tilde{S}^{[ij]}_{[\al\be]}] & =
&
S_L^2\tilde{\tilde{\De}}_{[0]}^{1}[\tilde{\tilde{U}}^i_{\al\,j}\,,\
\tilde{\tilde{S}}_{\al}^{[ij]}]\,.
\eeqa
As a matter of facts, this may be shown to result from previous constraints.

\subsubsection{$N_{d_{\al}} = 4$}

At this order, and using previous results, the cocycle condition writes :
\beq\label{i40}
S_L^2\De^2_{[0]} = 0 \ \ \Leftrightarrow \ \
S_L^2\tilde{\De}_{[0]}^2[\tilde{S}^{[ij]}_{[\al\be]}] = 0\,,
\eeq
which may be shown to result from the Ricci-flatness of $g'_{ij}$ ($\det\|g'\|$
= constant), which itself results from the HyperK\"ahler nature of the
perturbed metric $g'_{ij}\,,\ i.e.$ to results of the previous subsections.

\subsubsection{Stability of N=4 supersymmetric non-linear $\si$ models}

Putting the results of the previous subsections 5.2.1-5 altogether gives :
\beq\label{i50}
\De_{[0]} = \De_{[0]}^{an.(0)}[t_{(ij)}(\Phi)] +
\tilde{\tilde{\De}}_{[0]}^1 [\tilde{\tilde{U}}^i_{\al\,j}(\Phi),
\;\tilde{\tilde{S}}_{\al}^{[ij]}(\Phi)] -
\tilde{\De}_{[0]}^2[\tilde{S}^{[ij]}_{[\al\be]}(\Phi)]  +
S_L\De_{[-1]}[V^i(\Phi)]
\eeq
Up to a  trivial field and source reparametrisations (\ref{C7},\ref{C71}), the
most general Slavnov invariant integrated functional of the fields, sources and
their derivatives, of Faddeev-Popov 0 charge is :
\beqa\label{i51}
& \Ga^{'class.} & = \Ga^{class.} + \hbar \De_{[0]} \nnb\\
& \equiv & \int d^2x d^2\th \{g'_{ij}D_+\Phi^iD_-\Phi^j + \eta_i J_{A\,j}^{'i}[
d_A^{+}D_{+}\Phi^j + d_A^{-}D_{-}\Phi^j ] - {1\over2}\epsilon_{ABC}\eta_i\eta_j
J_{C}^{'ij}d_A^{+}d_B^{-}\}
\eeqa
and where the results of the previous subsections consistently give :
\beqa\label{i52}
g'_{ij}[\Phi] & = & g_{ij}[\Phi] + \hbar t_{(ij)}\ \ ,\ \ J_{3\,j}^{'i} =
J_{3\,j}^{i}\ \ ,\ \ J_{\al\,j}^{'i}(\Phi) = J_{\al\,j}^{i}(\Phi) + \hbar
\tilde{\tilde{U}}^i_{\al\,j}(\Phi)\ ,\nnb\\
J_{3}^{'ij}(\Phi) & = & J_{3\,k}^{'i}g^{'kj} = J_{3}^{ij} - \hbar J^i_n
g^{nm}t_{(mr)}g^{rj} \stackrel{\rm subsect.\ 5.2.3}{=} J_{3}^{ij}(\Phi)
+ 2\hbar \epsilon_{\al\be 3}\tilde{S}^{[ij]}_{[\al\be]}(\Phi)\ , \\
J_{\al}^{'ij}(\Phi) & = & J_{\al\,k}^{'i}g^{'kj} = J_{\al}^{ij}
+\hbar[\tilde{\tilde{U}}^i_{\al\,k}g^{kj} - J_{\al}^{im}t_{(mn)}g^{nj}]
\stackrel{\rm subsect.\ 5.2.2}{=} J_{\al}^{ij}(\Phi)
- 2\hbar \epsilon_{\al\be 3}\tilde{\tilde{S}}_{\be}^{[ij]}(\Phi)\ ,\nnb
\eeqa
and the different constraints (\ref{i11} and \ref{i21}) may be shown to enforce
N = 4 supersymmetry : $i.e.$ $g'_{ij}[\Phi]$ is a symmetric metric tensor, the
three $J'_A$ 's offer a set of anticommuting, integrable and covariantly
constant (with respect to the covariant derivative with connexion
$\Ga^{'i}_{jk}$ corresponding to the metric $g'_{ij}$) complex structures
satisfying a quaternionic multiplication law, and the metric is hermitian with
respect to each of these complex structure $J_{A\,j}^{'i}(\Phi).$

This proves the stability of the theory, and, thanks to the absence of anomaly,
the full renormalisability of N = 4 supersymmetric non-linear $\si$ models in
two space-time dimensions.

Of course, due to a possible Faddeev-Popov -1 charge non-trivial \footnote{\
The trivial cohomology $S_L\De_{[-1]}$ corresponds to field and source
reparametrisations according to (\ref{C7},\ref{C71}).} cohomology for $S_L$,
some of the parameters of the action (\ref{i51}) may be unphysical ones
. $\De_{[-1]}(V^i[\Phi])$ being independent of $d_{\al}^{\pm}$, the cocycle
condition $S_L\De_{[-1]}(V^i[\Phi]) = 0$ splits into the three ones
$S_L^i\De_{[-1]} (V^i[\Phi]) = 0\,,\ i=0,1,2$. One easily checks that
$V^i[\Phi]$ should be a contravariant Killing vector for the metric $g_{ij}$,
holomorphic with respect to the three complex structures :
\beq\label{i60}
g_{kj}\nabla_iV^k + g_{ki}\nabla_jV^k = 0\ ;\ \ J^{ij}_A\nabla_iV^k =
J^{ik}_A\nabla_iV^j\ , A=1,2,3.
\eeq

\section {Concluding remarks}

In the second paper of this series, we have analysed the cohomology of the
B.R.S. operator associated to N = 2 and N = 4 supersymmetry in a N = 1
superfield formalism. We found an anomaly candidate for torsionless models
built on compact K\"ahler Ricci-flat target spaces with a non vanishing Hodge
number  $h^{(3,0)}$. Calabi-Yau manifolds (3 complex dimensional case) where
$h^{(3,0)} = 1 $ (ref.\cite{2}) are interesting examples
due to their possible relevance for supertring theories. Of course, as no
explicit metric is at hand, one cannot compute the anomaly coefficient.

This anomaly in \underline{global} supersymmetry is a surprise with respect to
common wisdom \cite{18}. But some recent works of Brandt \cite{8} and Dixon
\cite{9} also show the existence of new non-trivial cohomologies in
supersymmetric theories and we have argued in Section 4 that the special
structure of the supersymmetry algebra which does not correspond to an ordinary
group but rather to a supergroup may be responsible of this peculiarity.

Our analysis then casts some doubts on the validity of the previous claims on
U.V. properties of N=2 supersymmetric non linear $\si$ models (see for example
\cite{3} or \cite{5}) : there, the possible occurence, at 4-loops order, of
(infinite) counterterms non-vanishing on-shell, even for K\"ahler Ricci-flat
manifolds, did not `` disturb" the complex structure. On the other hand, we
have found a possible `` instability" of the second supersymmetry, which
confirms that there are some difficulties in the regularisation of
supersymmetry by dimensional reduction assumed as well in explicit perturbative
calculations \cite{4} than in finiteness ``proofs" \cite{3} or higher order
counterterms analysis \cite{5}. We would like to emphasize the difference
between Faddeev-Popov 0 charge cohomology which describes the stability of the
classical action against radiative corrections ( the usual ``infinite"
counterterms) and which offers no surprise, and the anomaly sector which
describes the ``stability" of the symmetry ( the finite renormalisations which
are needed, in presence of a regularisation that does not respect the
symmetries of the theory, to restore the Ward identities) : of course, when at
a given perturbative order the Slavnov (or Ward) identities are spoiled, at the
next order, the analysis of the structure of the divergences is no longer under
control.
In particular, the Calabi-Yau uniqueness theorem for the metric \cite{19}
supposes that one stays in the same cohomology class for the K\"ahler form, a
fact which is not certain in the absence of a regularisation that respects the
N=2 supersymmetry (the possible anomaly we found expresses the impossibility to
find a regularisation that respects all the symmetries of these theories).

We emphasize that the present work relies heavily on a \underline{perturbative
analysis} of the possible breakings of the Slavnov identity, especially through
the use of the Quantum Action Principle. It may well happen that the
coefficient of the anomaly candidate vanishes at any finite perturbative order
\footnote{\ An argument based on the universality of the coefficient at any
finite order of perturbation theory and its  vanishing for a special class of
Calabi-Yau manifolds corresponding to orbifolds of tori, has been given to the
author by the referee of \cite{bo2}.}. However, the possibility of a
non-perturbative breaking of the N=2 supersymmetry would remain open.

Of course, if one has added from the very beginning extra geometrical (or
physical !) constraints that would fix the classical action, we bet that our
anomaly candidate would disapear : as previously mentionned, this is the case
when the manifold is a compact \underline{homogeneous} K\"ahler space.

Moreover we have been able to give the first algebraic, regularisation
free proof that, if one enforces N=4 supersymmetry (HyperK\"ahler manifolds)
there is no supersymmetry anomaly \footnote{\ This last result examplifies the
fact that non relevant cohomologies may appear in the filtration operation, at
first levels.} and that the corresponding non-linear $\si$ models are
all-orders renormalisable, ``\`a la Friedan".

The last step of our program will be the rigorous proof of the all-orders
finiteness of these models. We hope to be able to report on that subject in a
near future.

\bibliographystyle{plain}

\end{document}